\makeatletter
\def\input@path{{styles/}{styles/Imgs}{Sections/}{Figures/}}
\makeatother

\documentclass{article}

\usepackage{makecell}
\usepackage{multirow}
\usepackage{booktabs}
\usepackage{gensymb}
\usepackage{graphicx}
\usepackage{subcaption}
\usepackage[dvipsnames]{xcolor}
\usepackage{dsfont}
\PassOptionsToPackage{hidelinks,hyperfootnotes=false}{hyperref}
\usepackage{natbib}
\usepackage{doi}
\usepackage{amssymb}
\usepackage{amsmath}
\usepackage{pslatex} 
\usepackage{stmaryrd}
\usepackage{hyperref}
\usepackage[bottom]{footmisc}
\usepackage{styles/custom}
\usepackage{arxiv}

\date{}

\usepackage{authblk}

\newbox{\orcid}\sbox{\orcid}{\includegraphics[scale=0.06]{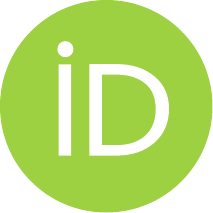}} 
\author[1,2]{\href{https://orcid.org/0009-0002-2151-8135}{\usebox{\orcid}\hspace{1mm}Philippe de Clermont Gallerande\thanks{\texttt{philippe.de-clermont-gallerande@inria.fr}}}}

\author[1]{\href{https://orcid.org/0000-0002-0101-3351}{\usebox{\orcid}\hspace{1mm}Quentin Avril\thanks{\texttt{quentin.avril@interdigital.com}}}}

\author[1]{\href{https://orcid.org/0000-0002-0973-4030}{\usebox{\orcid}\hspace{1mm}Philippe-Henri Gosselin\thanks{\texttt{philippehenri.gosselin@interdigital.com}}}}

\author[2]{\href{https://orcid.org/0000-0002-6160-8015}{\usebox{\orcid}\hspace{1mm}Ferran Argelaguet\thanks{\texttt{ferran.argelaguet@inria.fr}}}}

\author[2]{\href{https://orcid.org/0000-0002-7373-6049}{\usebox{\orcid}\hspace{1mm}Ludovic Hoyet\thanks{\texttt{ludovic.hoyet@inria.fr}}}}

\affil[1]{InterDigital R\&D France, Rennes, France}
\affil[2]{Inria, Univ. Rennes, CNRS, IRISA, Rennes, France}

\renewcommand{\shorttitle}{REACT}

\hypersetup{
pdftitle={REACT: Towards Real-Time Completion of Skeletal Motion Signals},
pdfsubject={Computer Science},
pdfauthor={Philippe de Clermont Gallerande, Quentin Avril, Philippe-Henri Gosselin, Ferran Argelaguet, Ludovic Hoyet},
pdfkeywords={Animation, Interactive Systems, Motion Capture, Neural Networks}
}

\begin{document}

\title{REACT: Towards Real-Time Completion of Skeletal Motion Signals}

\maketitle  \setcounter{footnote}{0}

\begin{minipage}{\linewidth}
    \centering
    \includegraphics[width=\linewidth]{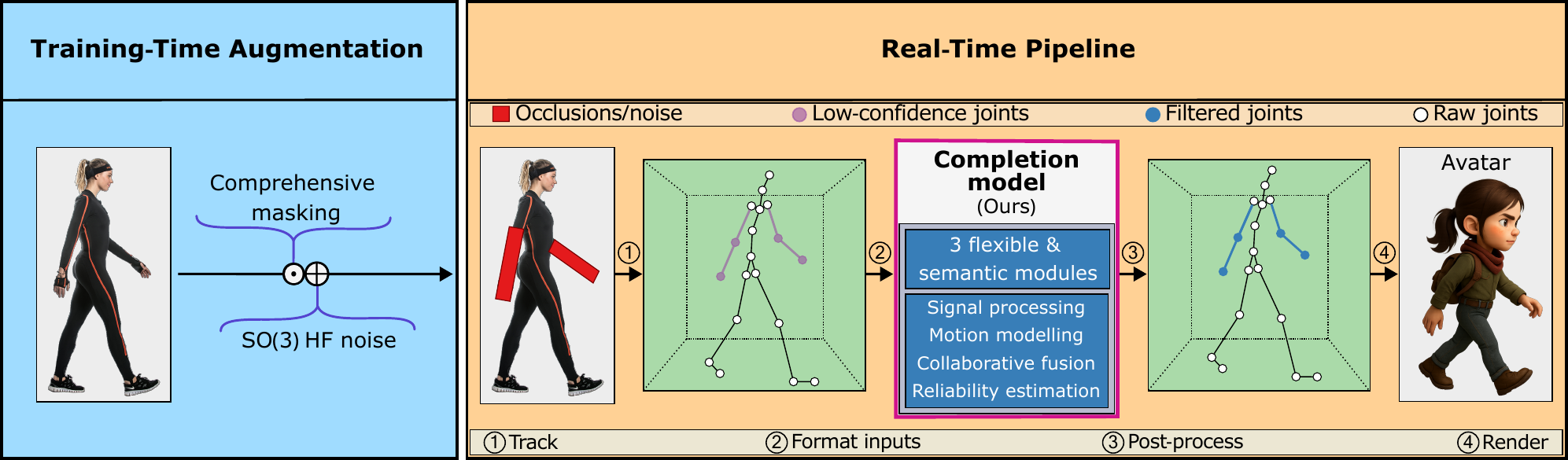}
    \captionof{figure}{REACT overview: training-time augmentation
    simulates joint occlusions and rotational noise; at inference,
    REACT detects unreliable joints and completes the streamed motion.}
    \label{fig:teaser}
\end{minipage}

\begin{abstract}
Tracking user motion in interactive and immersive applications is increasingly common (\eg, video games, XR systems, smartphones). Yet, processing and transmitting human motion data remain critical challenges, as current approaches strongly rely on the quality of the tracked motion signals. In general, affordable motion-capture systems may produce unnatural animations in the presence of calibration errors, occlusions, sensor drift, high-frequency noise, or other artifacts. More specifically, we simulate occlusions and high-frequency noise and propose to mitigate their effects with \emph{REACT}, an interactive online motion-completion framework operating causally, frame by frame, without look-ahead.
\emph{REACT} is an S4-based autoregressive model that coordinates complementary motion estimates under specific reliability-dependent constraints and exposes several interpretable internal signals. We present a system-level contribution consisting of a structured formulation of interactive motion completion together with a coherent instantiation of this formulation. We evaluate \emph{REACT} against several causal baselines and architectural variants under varying levels of occlusion severity, and report their respective inference latencies. Our results show that \emph{REACT} achieves a relevant compromise between reconstruction accuracy, update smoothness, and interactive inference. Finally, we characterize how its fusion behaviour changes under missing input observations.
\end{abstract}

\keywords{Animation, Interactive Systems, Motion Capture, Neural Networks}  
\section{Introduction}
In an increasingly connected world shaped by virtual communication channels, many media applications rely on advanced technologies to enhance users' expressiveness and sense of immersion. In particular, motion supports both exploration and non-verbal interaction with the environment, making its reliable modelling a crucial objective. Yet, its multimodal and highly contextual nature makes it difficult to model and reproduce, for instance when animating an avatar.

Dedicated motion-capture systems directly track and stream a user's body motion, shifting the animation problem from end-to-end conditional motion generation to the reconstruction and recovery of potentially corrupted signals. However, these technologies are expensive and often inconvenient: optical tracking typically requires multiple synchronized cameras, while IMU-based systems remain intrusive. They may also introduce characteristic noise into the input signal, such as IMU drift, or provide sparse observations by design, as in commercial-grade VR setups. Additional degradation may arise during online transmission or through storage-related transformations, such as format conversion or reduced-precision encoding. Even high-end capture pipelines commonly require time-consuming manual inspection and post-processing to detect and correct implausible poses~\citep{holden_optmocap_2018}.

In interactive systems, strict time constraints limit the available processing budget and therefore the possibility of carefully validating and correcting incoming signals, which may subsequently degrade the performance of downstream applications. Data-driven systems, whose behaviour may become unpredictable under out-of-distribution inputs, appear particularly vulnerable. Although motion-signal corruption has long been addressed in signal processing, computer graphics, and computer vision~\citep{MARTINI_filters_2025}, few existing methods jointly provide interactive operation, robustness to spatiotemporal sparsity (\ie, missing joint observations with varying spatial patterns and durations), and functional interpretability (\ie, the possibility of isolating functional parts of the pipeline to improve maintainability and auditability).

This combination of motion modelling, interactivity, robustness to spatiotemporal sparsity, and functional interpretability appears to remain underexplored in the literature. We address it through a structured deep-learning approach trained with a dedicated data-augmentation scheme for causal, interactive motion ``completion''. Standard motion inpainting and in-betweening methods reconstruct content within known temporal boundaries and generally exploit observations or keyframes located after the missing interval~\citep{hernandez_predinpaint_2019,starke_inbetweening_2023}, which limits their use in causal settings. Conversely, outpainting extends content beyond known boundaries and is conceptually closer to motion prediction~\citep{Wang_outpainting_2024}. In our frame-by-frame setting, neither future observations nor the duration of an ongoing corruption are known. We therefore use ``completion'' as an umbrella term encompassing filtering, causal recovery, and motion extrapolation. This formulation can be viewed as the reliability-aware coordination of several established motion-modelling tasks.

Our contributions are as follows:
\begin{itemize}
    \item We formulate streamed motion completion as the reliability-dependent coordination of three complementary motion sources: a directly transmitted observation, a reactive short-horizon prediction, and a longer-horizon generation.

    \item We instantiate this formulation in \emph{REACT}, an interactive, causal, and SO(3)-aware framework for completing streamed skeletal motion signals under \emph{a priori} unknown patterns of joint occlusion and noise severity. These degradations are simulated during training through a dedicated data-augmentation scheme. While most technical primitives are adapted from prior work, we either modify them or redesign their interactions to fit our setting. The model is organized into three functionally interpretable blocks. First, an adaptive input-denoising module combines classical filters through time-varying fusion. Second, a dual-head extrapolation backbone produces reactive short-horizon predictions and generates longer-horizon estimates. Third, a reliability-aware fusion module coordinates these estimates with direct observations using auxiliary cues derived from the filtering stage. The resulting correction is incorporated into a rolling buffer through an SO(3)-aware, PCAF-inspired update~\citep{barquero_rolling_2025} to improve temporal stability.

    \item We evaluate \emph{REACT} against several causal baselines and architectural variants using standard reconstruction and update-smoothness metrics, varying levels of occlusion severity, and inference-latency measurements. We further analyze its internal behaviour through the interpretable signals exposed by its structured design.
 
\end{itemize}
\section{Related Work}
As an umbrella term, motion completion covers a broad range of tasks. We refer interested readers to surveys focusing on motion filtering and completion~\citep{MARTINI_filters_2025}, as well as deep-learning-based approaches to human motion modelling~\citep{mourot_survey_2022,chen_character_2023,idrees_surveypred_2024}. In this section, we discuss motion prediction (Section~\ref{sec:Pred}), motion generation (Section~\ref{sec:Gen}), motion recovery (Section~\ref{sec:Recovery}), and human pose estimation from sparse trackers (Section~\ref{sec:PE}).

\subsection{Motion prediction}
\label{sec:Pred}
Motion prediction consists of extrapolating future motion from past observations. Closely related causal signal-estimation methods include classical filters such as the One-Euro filter~\citep{casiez_oef_2012} and the Kalman filter~\citep{kim_kalman_2018}. The One-Euro filter is an adaptive low-pass filter that balances smoothing and responsiveness according to signal velocity, but it is not designed for long-horizon prediction. By contrast, the Kalman filter explicitly estimates a future state from previous estimates and current observations. It therefore presents a useful structure for combining direct observations and predictions, although the standard linear formulation is poorly suited to highly nonlinear systems such as human motion.

Deep-learning approaches provide greater expressivity for modelling these nonlinear dynamics. In this paradigm, existing solutions can be broadly categorized as deterministic~\citep{DENG_detmotion_2024} or stochastic~\citep{LYU_motionpred_2022}. Within these categories, several architectural families have been explored, including recurrent models such as Long Short-Term Memory networks (LSTMs)~\citep{butepage_deep_2017} and Gated Recurrent Units (GRUs)~\citep{pavllo_quaternet_2018}, as well as Generative Adversarial Networks (GANs)~\citep{Barsoum_ganpred_2018} and Transformers~\citep{aksan_transfopred_2021}. Other distinctions arise from the problem formulation, such as sequence-to-sequence prediction~\citep{martinez_firstframe_2017} and autoregressive decoding~\citep{Zhang_autopred_2019}.

Several weaknesses of data-driven methods can be mitigated by injecting inductive biases~\citep{Zhang_PIMNet_2022}, for instance through domain-specific model structures. Zhang~\etal~\citep{Zhang_2024_WACV} propose to approximate an Euler--Lagrange-inspired formulation with a neural network and report strong and efficient short-term prediction performance ($<0.5$~s). This suggests that structured parametrizations may be particularly suitable for reactive short-horizon prediction.

In our setting, motion prediction is a natural approach for estimating the trajectory of temporarily missing joints. However, most prediction methods process the whole body rather than arbitrary subsets of missing joints. Furthermore, prediction ambiguity increases with the temporal horizon, making purely predictive approaches less reliable during extended occlusions. As the horizon increases ($>0.5~\text{s}$), the objective consequently becomes closer to producing a plausible continuation than to recovering the user's exact motion.

\subsection{Motion generation}
\label{sec:Gen}
Motion generation typically aims to produce plausible and diverse motion samples, including motions or transitions that may not be explicitly represented (or even absent) in the training data (\eg, gait transitions). Generative models are therefore often probabilistic and conditioned on control signals, with an associated sampling process.

Variational autoencoders (VAEs)~\citep{yan_mt-vae_2018}, which learn continuous latent spaces, became prominent for this task despite well-known failure modes such as \emph{mode averaging} and \emph{posterior collapse}. GANs~\citep{liu_aggregated_2021,malek-podjaski_adversarial_2022}, which rely on adversarial training, have also achieved strong performance, although they may suffer from \emph{mode collapse}. Architectures designed to better preserve the multimodal structure of motion have consequently gained traction. These include Vector-Quantized Variational Autoencoders~\citep{starke_codebook_2024}, based on discrete codebooks, and Normalizing Flows~\citep{henter_moglow_2020,yin_normflow_2021}, which learn bijective mappings between latent and data spaces. More recently, diffusion models~\citep{tevet_mdm_2023,karunratanakul_GMD_2023}, based on iterative denoising, have become prominent in motion generation. However, they are commonly applied to complete, fixed-size windows that are progressively refined, which raises specific challenges for causal and interactive use~\citep{xiao_motionstreamer_2025}.

The field relies on dedicated evaluations and objective metrics for measuring plausibility and diversity~\citep{Zhu_gensurvey_2024}. Since many applications involve human-oriented experiences, discrepancies between objective metrics and subjective perception remain an important concern~\citep{rekik_qa_2025}. More specifically, this generative perspective may be relevant to long-term occlusions: when the user's exact motion can no longer be predicted reliably, the remaining observed joints may instead serve as control signals for producing a plausible, diverse and perception-aware estimation of the whole-body motion.

\subsection{Motion recovery}
\label{sec:Recovery}
We use the term motion ``recovery'' to jointly refer to motion inpainting and in-betweening. These tasks complete a sequence along its temporal dimension (\ie, missing frames), its spatial dimension (\ie, missing joints), or both, while conditioning the result on available boundaries.

Typical motion-inpainting approaches include coarse-to-fine methods and sequence-to-sequence inference~\citep{he_inpaint_2023}. Hernandez~\etal~\citep{hernandez_predinpaint_2019}, for instance, exploit the proximity between motion inpainting and prediction by using a model trained for inpainting to perform future-motion prediction. Inpainting is also commonly applied to gaps in motion-capture data, although the resulting perceptual artifacts may not be adequately captured by objective metrics~\citep{kucherenko_mocap_2025}.

Motion in-betweening aims to smoothly connect two or more keyframes, often across relatively long temporal intervals, and is increasingly addressed through deep-learning methods~\citep{harvey_between_2020}. Beyond plausibility, human-interpretable and controllable generation is also desirable~\citep{cohan_diffinbetw_2024,starke_inbetweening_2023,starke_deepphase_2022}.

Standard inpainting and in-betweening methods nevertheless assume a well-delimited missing interval and generally exploit future observations or keyframes. They are therefore not directly applicable to our causal, frame-by-frame setting, in which neither the future pose nor the duration of an ongoing occlusion is known.

\subsection{Pose estimation from sparse signals}
\label{sec:PE}
Pose estimation from sparse observations has been studied for decades~\citep{chai2005}, but the growing adoption of immersive technologies and commodity motion-capture devices, such as commercial-grade Virtual Reality (VR) systems, has made it increasingly prominent. Interactive VR applications require per-frame inference, while the limited kinematic information available for each body region results in a severely under-constrained mapping from tracker signals to whole-body pose.

Strong motion priors are commonly used to reduce this ambiguity~\citep{yang_lobstr_2021,ye_neural3points_2022,jiang_avatarposer_2022,lee_questenvsim_2023}, although they may also limit expressivity and responsiveness to end-effector signals. Ponton~\etal~\citep{ponton_dragposer_2024} therefore propose an iterative optimization process within a structured latent space. Other approaches instead integrate pose estimation into a broader application-oriented framework, such as character control~\citep{starke_codebook_2024}. This can be achieved by aligning structured latent spaces for motion and control and processing them through a shared decoder.

In interactive applications, abrupt transitions caused by the disappearance or reappearance of an input signal may be particularly detrimental to user immersion. Classical filters may be insufficient because of their limited expressivity and adaptability. Barquero~\etal~\citep{barquero_rolling_2025} therefore combine a rolling buffer with a dedicated output parametrization of the model to encourage low-frequency smoothing while preserving a trade-off between reactivity and stability.

Most existing work nevertheless focuses on specific VR tracker configurations, which may limit generalization to arbitrary spatial and temporal patterns of missing observations. DragPoser~\citep{ponton_dragposer_2024}, for example, provides a flexible formulation but relies on iterative optimization, which may increase inference time for challenging samples. Our setting additionally requires the reliability of individual input signals to be estimated on-the-fly and without external information.

Taken together, these research directions motivate the functional structure of \emph{REACT}, presented in the next section. First, classical filtering provides auxiliary denoising estimates and reliability cues. Second, short-term motion prediction supports accurate motion extrapolation during brief occlusions. Third, motion generation offers a way to produce coherent and plausible motion when observations remain unavailable for longer periods. Finally, interactive, sparse pose-estimation methods motivate temporally stable update mechanisms, such as rolling buffers, as well as reliability-aware processing to better handle varying patterns of missing observations.

\section{Method}
\label{sec:Method}

Our main objective is to design a flexible framework for dynamically filtering and completing streamed motion signals under unknown spatiotemporal sparsity and noise, while preserving causal and interactive operation.

We first introduce the design requirements and provide a high-level overview in Section~\ref{sec:objectives}. We then detail the function and main characteristics of the different modules in Section~\ref{sec:archi}. Finally, Section~\ref{sec:opti} describes the training procedure, including the data-augmentation strategy.

\subsection{Objectives}
\label{sec:objectives}
To design a robust motion-completion method for streamed avatar animation, we impose several constraints on the task and framework. First, the method must operate online, or causally, without access to future observations, to remain interactive and responsive. This excludes approaches that rely on future frames, such as bidirectional sequence models. Second, inference must be sufficiently fast to support interactive operation. Standard diffusion-based methods and iterative optimization approaches are therefore difficult to apply without substantial adaptation. Third, the system must adapt its per-joint behaviour according to the estimated reliability of each input measurement, which is unknown \emph{a priori}.
Finally, we focus on joint rotations within a fixed skeletal topology. This choice isolates motion completion from morphology and retargeting variations, while facilitating rotation-aware interpolation and correction.

\subsection{Architecture}
\label{sec:archi}
The research directions discussed in the previous section motivate three functionally distinct modules. First, an input-filtering module combines classical filters to reduce high-frequency noise and limit unnatural jerkiness. It also provides filtered estimates, residuals, and other auxiliary cues subsequently used for reliability estimation. Second, an S4-based backbone~\citep{gu_s4_2022} produces two complementary completion candidates: a reactive short-horizon prediction (denoted \emph{Phys}) and a plausible longer-horizon estimate (denoted \emph{Data}). Third, a reliability-aware fusion module coordinates these two candidates with a direct observation path (denoted \emph{Ref}). In addition, for each joint, the fusion module estimates input reliability, adjusts the contribution of the three sources according to the current context, and improves temporal and geometric stability through an SO(3)-aware, PCAF-inspired rolling update~\citep{barquero_rolling_2025}.

Before detailing these modules, we describe the input buffers, model outputs, and recurrent states used by the complete framework.

\begin{figure}[t]
    \centering
    \includegraphics[width=1\linewidth]{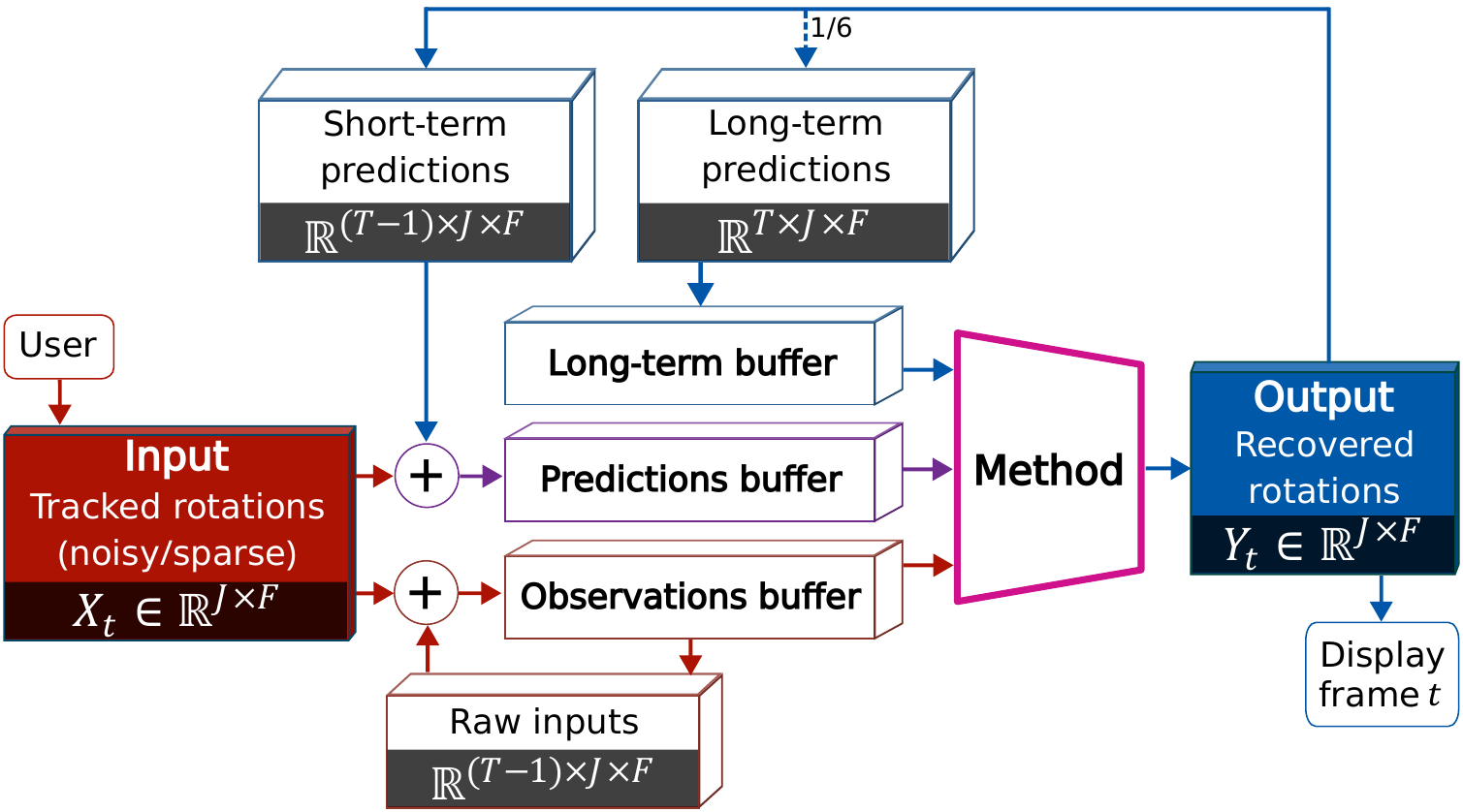}
    \caption{Our system takes as input several history buffers. At each step, the current noisy/occluded frame is stacked with $T-1$ past predictions and $T-1$ past observations. This explicit short context is used by the network together with a longer temporal horizon, provided by a third buffer, which samples model predictions at a lower frequency.}
    \label{fig:overview}
\end{figure}

\para{Inputs \& outputs}
\label{subsec:insandouts}
As illustrated in Figure~\ref{fig:overview}, our approach receives a frame $X_{t} \in \mathbb{R}^{J \times F}$ containing missing or noisy measurements, where $J$ is the number of joints of the skeleton and $F$ is the joint rotation representation dimension. Across time, we store three sequences of past frames with distinct semantic meanings.
First, the \textit{Observations buffer} $O_t$ stacks raw inputs, \ie, $ O_t=\{ X_{t-T+1}, \dots, X_{t} \} \in \mathbb{R}^{T \times J \times F} $.
Second, the \textit{Predictions buffer} $P_t$ contains the previous model outputs together with the current observation. Let $Y_t\in\mathbb{R}^{J\times F}$ denote the completed pose at frame~$t$. To prevent information leakage during training, the current entry of $P_t$ is $X_t$ rather than $Y_t$: $P_t=\{Y_{t-T+1},\dots,Y_{t-1},X_t\}\in\mathbb{R}^{T\times J\times F}$.
Third, the \textit{Long-term buffer} $L_t$ provides a coarser temporal context by uniformly subsampling past predictions at one sixth of the frame rate, with periodic phase alignment. Let $\delta = 1+ (t \bmod 6)$ denote the phase, then $L_t = \{ Y_{t-(6k+\delta)} \mid k\in \llbracket 0,T-1 \rrbracket \} \in \mathbb{R}^{T \times J \times F}$.
Our input is then given to the model and processed to provide a more reliable frame at time~$t$. Importantly, beyond the current input, our method maintains several persistent states and history buffers across inference steps. In practice, we use $T = 31$ to have $\approx 0.5\,\text{s}$ of short temporal context at $60$~Hz, $J = 23$ and $F=6$~\citep{zhou_continuity_2019}. The long-term buffer has the same temporal dimension but a lower sampling frequency, which provides a broader temporal context for human motion. In practice, it stores $31$ frames at $10$~Hz, which corresponds to $\approx 3~\text{s}$ of memory.

\para{S4 \& dual-head output}
\label{subsec:S4}
State-space models were popularized by the Structured State Space Sequence (S4) architecture~\citep{gu_s4_2022} and are ruled by $\dot x = Ax + Bu \text{ and } y = Cx + Du$, where $x$ denotes the internal state and $u$ the input signal.

\begin{figure}
    \centering
    \includegraphics[width=0.95\linewidth]{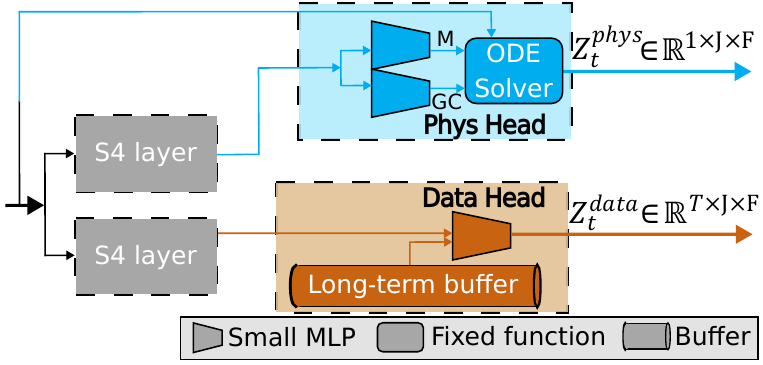}
    \caption{S4-based dual-branch backbone. A shared input feature sequence is processed by two specialized S4 branches. The \emph{Phys} branch, depicted in light blue, uses an ODE-inspired parameterization~\citep{Zhang_2024_WACV} to produce reactive and precise short-horizon predictions. The \emph{Data} branch, depicted in orange, uses longer-term context to regress a plausible future motion.}
    \label{fig:2S4}
\end{figure}

In our model, a shared embedding is processed by two specialized branches (\cf~Figure~\ref{fig:2S4}). The split between these branches may occur at different levels: each branch may use an independent S4 layer, or both may rely on a single partitioned, dual-headed S4 layer. We focus here on the \texttt{-2S4} variant, which yielded better preliminary results, while alternative configurations are described in the supplementary material. The two branches are then mapped to distinct functional outputs. The \emph{Phys} branch uses an ODE-inspired parameterization to produce reactive short-horizon predictions, whereas the \emph{Data} branch uses longer-term context to regress a plausible future trajectory.

\subpara{Physics-inspired head (``Phys'')}
The first branch performs a pure motion prediction. To prevent information leakage from the potentially corrupted current observation, it discards the current frame and estimates the pose at time~$t$ from the preceding motion context. To introduce a suitable short-term inductive bias, we adapt the output parametrization proposed by Zhang~\etal~\citep{Zhang_2024_WACV}.

At a high level, the branch learns quantities involved in the generalized Euler--Lagrange equation $M\ddot{q} + C_{\mathrm{bias}} = G$, where $M$ denotes the generalized inertia matrix, $C_{\mathrm{bias}}$ represents generalized bias forces, including Coriolis and gravitational effects, $G$ represents the remaining generalized forces, such as ground-reaction forces, and $\ddot{q}$ is the second-order temporal derivative of the pose parameterization. More specifically, the branch estimates $M$, $C_{\mathrm{bias}}$, and $G$ from the available temporal context, and derives an acceleration estimate as $\ddot{q} = M^{-1}(G-C_{\mathrm{bias}})$. The current pose representation is then extrapolated using a second-order finite-difference approximation:
$\ddot q_t=\frac{q_t-2q_{t-1}+q_{t-2}}{\Delta t^2},\ q_t=\ddot q_t\Delta t^2+2q_{t-1}-q_{t-2}$, where $\Delta t=1/60$~s.

In practice, $M_t$ is constructed at runtime through a Cholesky factorization, ensuring that it is symmetric and positive definite. Rather than estimating $G$ and $C_{\mathrm{bias}}$ separately, we directly regress their difference with a single network, denoted $GC$ in Figure~\ref{fig:2S4}, as separating the two terms did not provide an apparent benefit in our preliminary experiments. These quantities are estimated in the 6D joint-rotation representation space.

This design should be understood as a physics-inspired inductive bias rather than a physical simulator. In particular, the estimated quantities should not be interpreted as physically meaningful inertia or force terms. The branch neither recovers physical forces nor enforces valid full-body dynamics; instead, it constrains local motion evolution according to the recent context and provides a reactive short-horizon completion candidate.

As autoregressive extrapolation becomes increasingly ambiguous and may accumulate errors over longer horizons, the second branch provides a complementary longer-term estimate.

\subpara{Data head (``Data'')}
The \emph{Data} branch aims to produce stable motion trajectories using the broader temporal context provided by the long-term buffer. It is the only component of the network that explicitly estimates multiple future poses at each time step. The first pose of the predicted sequence directly contributes to the completion of the current frame, while the wider prediction horizon encourages a coherent continuation when observations remain unavailable for extended periods.
At longer horizons, recovering the user's exact motion becomes inherently ambiguous. Accordingly, the role of this branch is not to reconstruct a unique ground-truth trajectory, but to provide the downstream fusion module with a smooth and plausible motion candidate conditioned on the available observations.

Overall, the backbone produces two complementary completion candidates. The \emph{Phys} branch provides a reactive extrapolation suited to brief signal losses, whereas the \emph{Data} branch provides a more stable extrapolation over longer intervals. Both branches regress joint rotations in the continuous 6D representation, and their outputs are independently orthonormalized~\citep{zhou_continuity_2019} to facilitate their future interpretation as rotation matrices. For brevity, we use \emph{Phys} and \emph{Data} to refer either to the corresponding branch or to its output candidate when the distinction is unambiguous.

\para{Fixed input filters}
\label{subsec:in_filt}
The core S4 layer is a Linear Time-Invariant (LTI) system, which means its state matrices remain static at inference. Thus, the filter does not adapt to local features of the signal, such as a higher energy in high-frequency components. To illustrate the issue in a motion scenario, a filter with a low cut-off frequency is relevant to denoise locomotion, but would fail to capture salient high frequency motions such as waving a hand. Furthermore, it would likely distort the signal if there were changes in the velocity, such as going from walking to jogging to running. Conversely, a filter with a higher cut-off frequency would capture these changes but would not filter noise as efficiently. 
This limitation is addressed by selective state-space models such as
Mamba and Mamba-2~\citep{dao_mamba2_2024}, whose state parameters
depend on the input and can therefore adapt their dynamics to the local
signal content. However, as these architectures are both more complex to modify and to train than the original S4, we instead adopt an explicit filtering module.

Our filtering module relies primarily on three small and efficient filters commonly used for streamed signals. First, we use an alpha-beta-gamma tracking filter as a lightweight smoothing mechanism for high-frequency input noise. Second, we apply a causal five-frame moving-median filter to suppress isolated single-frame anomalies. Finally, we use the One-Euro filter~\citep{casiez_oef_2012}, which is specifically designed to balance smoothing and responsiveness in interactive signals. To combine their outputs, we use a learned gating mechanism: an MLP extracts local signal features and regresses time-varying fusion logits. Then, a combination of raw and
filtered signals is provided to the backbone. 
In parallel, we construct an auxiliary context signal, which include several filter residuals and the filter-mixture weights produced by the gating mechanism. This signal is later used by the fusion module for per-joint reliability estimation.

\para{Collaborative fusion}
\label{subsec:fusion}
At the output of the backbone, the system has access to three complementary completion candidates. \emph{Phys} and \emph{Data} are produced by their corresponding predictive branches, while a skip-connection directly transmits the current input observation, which we denote \emph{Ref} (for reference). When reliable, \emph{Ref} provides an anchor that preserves the measured motion. Its discrepancies with the predictive candidates also provide auxiliary cues about possible input corruption.

These three candidates correspond to complementary operating regimes. First, when a joint observation is reliable, the system should intervene as little as possible to avoid introducing model-induced distortions, such as damping or overshooting. Second, observations may become unavailable for a brief period because of a short self-occlusion, an intermittent tracking failure, or a transient signal artifact. In this regime, \emph{Phys} is a reactive candidate constrained by the recent motion context. Its disagreement with \emph{Ref} may additionally constitute a useful reliability cue. Third, when an observation remains unavailable for a longer period, \emph{Ref} becomes uninformative and the autoregressive \emph{Phys} estimate may accumulate errors. The \emph{Data} candidate then provides a more stable and plausible continuation.

The per-joint reliability of the observations is therefore central to the coordination of these regimes and must be estimated online. Because unconstrained coordination between parallel branches with distinct functional roles may lead to undesirable solutions, we structure the fusion around this estimated reliability and introduce several inductive biases during training.

For clarity, we separate the fusion module into four main operations, described below.

\subpara{Aggregation weights}
An MLP predicts the interpolation weights used to combine the three completion candidates. Its inputs include: (i) the \emph{Ref}, \emph{Phys}, and \emph{Data} candidates and their temporal differences; (ii) the angular differences between the predictive candidates; and (iii) learned projections of the long-term buffer and S4 states. The MLP is optimized indirectly through the motion objectives,  together with several regularization terms whose coefficients are annealed towards zero during training. These include entropy maximization, an early-stage penalty encouraging balanced weights, and behaviour-specific penalties. In particular, persistent low reliability progressively discourages the use of \emph{Ref} and, over longer intervals, \emph{Phys}, in favour of \emph{Data}.

Recall that \emph{Phys} and \emph{Ref} each provide a single-frame candidate, whereas \emph{Data} produces a sequence. Since the fused output must also form a sequence before being incorporated into the rolling buffer, we repeat the \emph{Phys} and \emph{Ref} candidates along the temporal dimension, corresponding to a zero-velocity extension. Consequently, \emph{Data} is the only candidate that explicitly proposes distinct corrections for distant future frames.

The aggregation MLP outputs $Z_{\mathrm{agg}}\in\mathbb{R}^{T\times J\times4}$. For each frame and joint, the four logits are divided into two pairs and normalized independently through two softmax operations. The candidates are then combined through two staged one-versus-one interpolations in the quaternion space. For example, interpolating \emph{Phys} and \emph{Data} first, and then comparing the result with \emph{Ref}, yields 
$Y=\beta\,Y_{\mathrm{Ref}}+(1-\beta)\left(\alpha\,Y_{\mathrm{Phys}}+(1-\alpha)\,Y_{\mathrm{Data}}\right)$, where $\alpha,\beta\in[0,1]$. The corresponding effective weights are therefore $w_{\mathrm{Ref}}=\beta, w_{\mathrm{Phys}}=(1-\beta)\alpha, w_{\mathrm{Data}}=(1-\beta)(1-\alpha)$.
This hierarchical formulation factorizes the fusion into two interpretable pairwise decisions. In particular, the candidate introduced at the second stage receives a direct gate against the result of the first interpolation.

Different interpolation orders therefore define distinct structural configurations and can affect performance. For example, placing \emph{Ref} in the second interpolation stage gives it a direct gate against the combined predictive estimate, which may be appropriate when observations are expected to be reliable most of the time.

\subpara{Input-signal reliability}
Because reliability estimation is central to the fusion process, we use two complementary mechanisms. First, a reactive frame-wise classifier predicts a per-joint unreliability probability. It is trained with a binary cross-entropy loss against the ground-truth occlusion masks used during data augmentation. It receives: (i) the angular differences between \emph{Phys} and \emph{Data}; (ii) the filtering-module outputs, including residuals and mixing weights; and (iii) a detached copy of the history buffer described next.

Second, we maintain a persistent per-joint state, referred to as the \emph{defiance buffer}, which acts both as temporal memory and hysteresis. Broadly, this buffer accumulates evidence of unreliability over time, so that a large value indicates that the joint has been considered unreliable over several successive frames. The amplitude of its updates further depends on the confidence of the instantaneous classifier. This mechanism stabilizes the reliability estimate and reduces abrupt behavioural changes, which is desirable in the presence of frame-wise misclassifications (\ie, false negatives/false positives) or blinking occlusions.

\subpara{Aggregation priors}
We additionally inject explicit priors into the aggregation weights according to the estimated operating regime and the temporal position within the predicted sequence. For example, \emph{Data} is progressively favoured for distant sequence entries because \emph{Phys} and \emph{Ref} are extended through zero-velocity repetition rather than explicit future prediction.

These priors combine hard-constrained exponential decays parameterized by half-lives with small dynamic corrections regressed by MLPs. This design preserves the intended temporal behaviour while allowing limited context-dependent adaptation.

We also modulate the temperature of the pairwise softmax operations as a function of the geodesic disagreement between the completion candidates. The fusion can therefore make sharper or smoother decisions depending on how strongly the candidate rotations disagree.

\subpara{Correction amplitude bottleneck}
The fourth and final block builds on a variant of the Prediction Consistency Anchor Function (PCAF) proposed by Barquero~\etal~\citep{barquero_rolling_2025}. This mechanism reparametrizes the network output as a correction applied to a rolling buffer. It therefore limits the influence of the current network prediction $f_{\theta}(X_t, C_t)$ on the immediately emitted frame and promotes smoother motion trajectories. In the original work, PCAF is formulated as $P_t = P_{t-1} + U \cdot \tanh\!\left(f_\theta (X_t, C_t) - P_{t-1}\right)$, where $U$ denotes an uncertainty-dependent weighting function along time.

We introduce two modifications to this reparametrization. First, the maximum correction amplitude is dynamically adjusted using learned parameters and a defiance signal, which combines the normalized defiance buffer, predicted occlusion probabilities, and inter-branch differences. This modulation may increase or decrease the correction cap, or entirely bypass the amplitude bottleneck when defiance is zero. This mechanism aims to limit over-smoothing and facilitate training.

Second, we reformulate the correction on SO(3). The relative rotation is mapped to the tangent space $\mathfrak{so}(3)$, where its direction is preserved while its geodesic magnitude is scaled and bounded according to:

\begin{equation}
    \theta_{\mathrm{out}} =
    U\theta_{\mathrm{cap}}
    \tanh\!\left(
        \frac{\theta_{\mathrm{in}}}{\theta_{\mathrm{cap}}}
    \right),
\end{equation} 
where $\theta_{\mathrm{in}}$ and $\theta_{\mathrm{out}}$ denote the input and corrected rotation angles, respectively. This formulation also enables interpretable per-joint constraints on the correction amplitude based on contextual and interpretable signals, such as the recent velocity history.

Finally, the resulting corrections are incorporated into the rolling buffer through rotation-matrix composition. Its first entry is then removed and emitted as the completed frame. The emitted frame is also copied to initialize the newly appended tail entry.

\subsection{Optimization}
\label{sec:opti}
The training pipeline comprises several stages intended to improve robustness to missing and noisy observations.

\para{Masking}
We first introduce a data-augmentation scheme designed to approximate the signal degradations considered in our completion task. Following Cormier~\etal~\citep{Cormier_SkelDA_2024}, we stochastically mask random numbers of joints over temporal intervals of varying duration. During training, selected joint signals are either (i) corrupted with rotation-aware Gaussian noise or (ii) replaced with a masking value to simulate an unavailable observation. Although analogous to the masking tokens used in language modelling~\citep{devlin_bert_2019}, our masking value is not a learned embedding. Instead, it is sampled from a narrow distribution centred on the identity rotation.

The masking procedure is controlled by several hyperparameters. In particular, $M_{\min}$ and $M_{\max}$ define the minimum and maximum target proportions of masked joints. For each sample, a continuous target ratio is sampled as $m\sim\mathcal{U}(M_{\min},M_{\max})$ and converted into an integer target number of masked joints by rounding to the nearest integer. This discretization assigns non-zero probability to target counts of zero masked joints when $M_{\min}=0$ and all $J$ joints when $M_{\max}=1$. In our experiments, $M_{\min}=0$, while $M_{\max}$ ranges from $25\%$ to $100\%$. The expected continuous target ratio is
therefore $\mathbb{E}[m]=(M_{\max}+M_{\min})/2$.

The realized proportion of masked joints may nevertheless differ from $m$, because the spatial masking procedure can select complete anatomical regions rather than individual joints. With probability $60\%$, joints are selected according to anatomical regions; otherwise, they are selected independently.

Masked joints are temporally persistent rather than independently resampled at every frame. Each activation is associated with a random duration, with both short- and longer-duration regimes. Existing masks may be extended when overlapping joint- or body-part masks are sampled, and additional short bursts may occur around mask expiration. Consequently, the effective duration of a joint occlusion depends both on its sampled lifetime and on subsequent mask renewals. The target masking ratio controls the number of newly activated joints, but does not uniquely determine their realized duration. Overall, this procedure introduces temporal persistence while avoiding fixed or easily predictable occlusion patterns.

\para{Autoregression}
We use scheduled sampling~\citep{bengio_scheduler_2015} to mitigate exposure bias by gradually increasing the probability of feeding the network its own previous outputs during training. In addition, we use a free-running rollout strategy similar to that proposed by Barquero~\etal~\citep{barquero_rolling_2025}, exposing the model to autoregressive sequences of up to 600 frames. Finally, sequences within each batch are padded to the length of the longest sequence using the identity rotation, and the corresponding padded elements are excluded from the loss computation.

\para{Branches balancing}
Short-term prediction is intuitively easier than longer-horizon forecasting. In addition, the stronger output constraints imposed on the \emph{Phys} branch generally limit its initial prediction errors. Without explicit balancing, the fusion consequently tends to collapse towards \emph{Phys}, depriving the \emph{Data} branch of informative gradients. Conversely, unrestricted end-to-end optimization may cause \emph{Phys} to drift away from the reactive short-term behaviour learned during its initial specialization stage.

We therefore train the model in three stages. First, the \emph{Phys} branch is trained independently on short-term prediction, with autoregressive rollouts limited to at most 10 consecutive frames. At the end of this stage, we save a snapshot of its parameters.
Second, the \emph{Data} branch is trained independently using longer and more frequent autoregressive constraints. Because this branch predicts a sequence of poses, its loss supervises all frames of the predicted sequence rather than only the current-frame candidate.
Third, the complete model is optimized end to end. We introduce three mechanisms to preserve the specialization of the branches and prevent fusion collapse. First, each batch contains a proportion of samples processed under teacher forcing, either with clean observations or with additional noise. Second, we apply an $\ell_2$-SP regularization term~\citep{li_l2sp_2018} that penalizes deviations of the \emph{Phys} parameters from the snapshot obtained after the first stage. Third, during the early stages of joint training, we prevent either pairwise fusion decision from saturating by applying an $\ell_2$ penalty that pulls each binary gate towards $0.5$. This does not impose uniform effective weights over the three candidates, since the hierarchical interpolation remains structurally asymmetric. Rather, it initially keeps both alternatives active at each fusion stage,before the regularization coefficient is annealed towards zero.
Entropy maximization and the expected-behaviour penalties introduced previously provide additional guidance during training. These terms, collectively denoted $L_{regu}$, use small coefficients, with a maximum magnitude of approximately $0.001$, and selected coefficients are progressively annealed towards zero.

\para{Distillation}
Finally, we investigate whether the complete structured model can be compressed through knowledge distillation~\citep{hinton_distil_2015}. During training, a frozen teacher and a smaller student process the same corrupted observations, but maintain their own prediction buffers and autoregressive states. The student is initially trained mostly under teacher forcing, after which the probability of reusing its own previous outputs is progressively increased through scheduled sampling.

Unlike the complete model, the student does not maintain a rolling correction buffer and directly emits one completed frame at each step. It is optimized using the same ground-truth motion objectives as the original model, together with an additional distillation term penalizing the geodesic discrepancy between the frame emitted by the student and the corresponding teacher output.

The student is therefore encouraged both to reconstruct the ground-truth motion and to reproduce the output behaviour of the structured teacher, while learning to remain stable under its own autoregressive predictions. Our objective is to obtain a compact model offering a more favourable trade-off between motion quality, inference latency, and memory footprint. The student architectures and corresponding results are presented in the following section.
\section{Experiments}
\label{sec:Experiments}
\subsection{Details}

\para{Loss and optimizer}
\label{subsubsec:hyperp}
In addition to the regularization terms specific to our method ($L_{regu}$), we use a $\ell_1$ ($w=1e^{-6}$) and a weight decay ($w=0.01$) on the learnable parameters. Note that we do not apply these two regularizations directly on the S4 state matrices, as advised in the official implementation. The general loss of the task involves several differentiable criteria, such as MPJPE ($w=1$), MPJAE ($w=0.1$), MPJLAE ($w=0.01$), MPJPVE ($w=0.001$) and MPJAVE ($w=0.001$), which we aggregate as $L_{rec}$. For more details about the metrics definition, please refer to the supplementary materials. Broadly, the MPJPE is a per-joint positional error, the MPJAE and MPJLAE are both per-joint rotational errors (respectively in world and local coordinates), and the MPJPVE and MPJAVE are the errors on the first order time derivatives of, respectively, the positional and angular parameters. We further include a binary cross-entropy term denoted $L_{bce}$ ($w=0.1$) to train the occlusion classifier, and a Frobenius distance between the teacher's and the student's output rotations ($w=1$) for the distillation, when relevant.
Note that MPJ* criteria are necessarily averaged over joints and may be computed after a forward kinematics when relevant. 

The overall loss for REACT is then:
\begin{equation}
\mathcal{L}= \frac{1}{TJ} \sum_{t=0}^{T-1}\sum_{j=0}^{J-1} \left[ L_{rec}(\hat{R}_{j,t},R_{j,t}) + L_{bce}(\hat{m}_{j,t},m_{j,t}) \right] + L_{regu}(\theta)
\end{equation}
with $\hat{R}$ and $R$ the predicted and ground-truth rotation matrices, $\hat{m}$ and $m$ the predicted and ground-truth masks, $\theta$ the learnable weights, $J$ the number of joints and $T$ the number of frames.

Models are trained for 17 epochs with batch size 32 using the AdamW optimizer, an initial learning rate of ($1e^{-4}$), halved every four epochs, and a per-step triangular cyclical scheduler.

\para{Evaluated models}
\label{subsubsec:models}
We evaluate causal baselines that receive both the prediction and observation buffers. These include a modified version of the Rolling Prediction Model~\citep{barquero_rolling_2025} (\texttt{RPM}) and two GRU-based baselines, \texttt{GRU-ctx} and \texttt{GRU-ctx-large}, which share the same overall architecture but differ in depth and width. The \texttt{-ctx} suffix indicates that these models receive the same two context buffers as \texttt{RPM}.

The original \texttt{RPM} targets online filtering for XR settings in which the sparse tracker configuration is known and stable. Its pipeline therefore includes assumptions and heuristics tailored to specific tracker layouts, such as hand controllers. These assumptions do not hold in our setting, where the spatial pattern and reliability of the observations vary over time. When directly applied to our task, the original model converges toward small or null corrections, producing excessively stable and nearly static pose sequences.

We therefore introduce two adaptations. First, we replace its sparse tracker buffer with our dense raw-observation buffer. Second, we use the known synthetic corruption mask to blend the PCAF output with a less constrained correction through an exponential heuristic. This allows larger corrections when a joint is observed and mitigates the slow-update behaviour induced by PCAF. Consequently, this adapted \texttt{RPM} has access to the ground-truth reliability mask, which is
not available to \emph{REACT} at runtime.

We additionally evaluate two distilled students: a VQ-VAE-based student (\texttt{VQVAE-ctx-student}) and \texttt{GRU-ctx-student}, which uses the same architecture as \texttt{GRU-ctx}. These baselines are compared with three variants of our method. We discuss additional architectural configurations of our method in
the supplementary material.

\emph{REACT} uses an S4-based dual-output backbone framed by the filtering and fusion modules described previously. The \texttt{-learned} suffix denotes the learned gate used to combine the filtered input signals, while \texttt{-2S4} denotes a backbone with one S4 layer per predictive branch. Finally, the suffixes \texttt{-r}, \texttt{-p}, and \texttt{-d} identify the candidate introduced at the second interpolation stage: \emph{Ref}, \emph{Phys}, and \emph{Data}, respectively.

\para{Dataset}
\label{subsubsec:dataset}
For our experiments, we use the CMU dataset preprocessed by Holden~\etal~\citep{holden_deep_2016}. This choice is motivated by its size, its widespread use in the literature, and the diversity of covered motions. The resulting dataset contains $2,423$ animations ($2,082,000$ frames at 60~Hz), including actions such as walking, running, dancing, boxing, and digging. We split the dataset at the animation level, allocating respectively $50\%$, $20\%$, and $30\%$ of the dataset to training, validation, and testing. At each epoch, approximately $15\%$ of the training set is randomly sampled rather than processing the complete set. Finally, the skeleton topology and morphology are fixed and normalized, allowing us to focus on temporal and rotational motion modelling.

\subsection{Metrics}
Among our reported metrics, we include positional errors such as MPJPE (in world coordinates) and Procrustes-aligned MPJPE (P-MPJPE). The latter reduces the influence of global misalignment on the resulting pose error. We also measure joint-rotation errors through geodesic distance, applied to both local rotations (MPJLAE) and global rotations (MPJAE). Although positional metrics implicitly encode part of the rotational information, they cannot reliably assess errors across all rotational degrees of freedom and may inadequately capture errors on distal joints~\citep{kucherenko_mocap_2025}.

To evaluate temporal stability during changes of operating regime, we use the Peak Jerk (PJ) and Area Under the Jerk (AUJ) metrics introduced by Barquero~\etal~\citep{barquero_auj_2024}. Following their work, PJ and AUJ are reported separately for the two update regimes. We use the term ``update regime'' rather than ``transition'' to emphasize the behavioural shift between prediction and tracking. In particular, ${S\to T}$ (Synthesis to Tracking) corresponds to de-occlusion, whereas ${T\to S}$ (Tracking to Synthesis) corresponds to the onset of an occlusion. Since these metrics remain insufficient to characterize whether the models preserve the broader structure of human motion, we discuss additional complementary metrics in the supplementary material.

\subsection{Quantitative results}
\para{Complexity}
We measure several aspects of model complexity using PyTorch profiling tools and synchronized CUDA events. Each inference measurement includes feature encoding and decoding as well as all recurrent buffer and state updates. We additionally report the maximum VRAM reserved by PyTorch during inference. Results are summarized in Table~\ref{tab:complexity}.

\emph{REACT} has an intermediate parameter count compared with the evaluated baselines, but incurs substantially higher inference latency and memory requirements. On the evaluated RTX~3090, its measured latency remains compatible with interactive operation, at approximately $22$--$24$~ms per frame, although it does not reach the 60 or 90~Hz rates commonly targeted by immersive applications. By contrast, both distilled students operate comfortably above 500~fps while requiring a substantially smaller memory footprint.

\begin{table}[ht!]
\caption{Complexity and runtime. Hardware: RTX~3090 (24 GB), B=1, T=31, J=23, precision FP32. Environment: PyTorch 2.3.1, CUDA 12.1, cuDNN 8.9.2. Metrics are aggregated over 500 iterations following 100 steps of GPU warm-up. We further report the 50th and 95th percentiles (P50 and P95).}
\label{tab:complexity}

\centering
\small
\setlength{\tabcolsep}{2pt}
\begin{tabular}{lccc}
\toprule
Model & \makecell[c]{Params [M]} & 
\makecell[c]{Latency [ms] \\ P50 | P95}  &
\makecell[c]{VRAM [MB]} \\
\midrule
\midrule
RPM   & 6.75 & 4.03 | 4.30 & 98 \\
\midrule
GRU-ctx & 0.38 & 0.57 | 0.64  & 24\\
GRU-ctx-large & 8.09 & 1.41 | 1.60 & 148\\
\midrule
VQVAE-ctx-student & 0.38 & 1.47 | 1.78 & 24 \\
GRU-ctx-student & 0.38 & 0.62 | 0.66 & 22 \\
\midrule
REACT-learned-2S4 & 1.10 & 21.94 | 23.71 & 222 \\
\bottomrule
\end{tabular}
\end{table}

\para{Completion}
\label{subsubsec:completion}
Our primary evaluation assesses motion completion against the ground-truth signal. We first compare our method with the baselines under a fixed training and evaluation configuration, before studying its sensitivity to changes in occlusion severity.

\subpara{Completion metrics}
Tables~\ref{tab:fra_met} and~\ref{tab:seq_met} report the quantitative results of the main evaluation, with all models evaluated under the same masking configuration and results reported as $\mathrm{mean}\pm\mathrm{std}$.

On pose-level metrics, all three \emph{REACT} variants outperform the evaluated baselines, although \texttt{GRU-ctx} narrows the gap on positional errors. The largest improvements are observed on rotational metrics, with \texttt{REACT-learned-2S4-r} reaching the lowest MPJLAE and MPJAE. The distilled students also substantially reduce rotational errors compared with the GRU baselines, but this improvement does not translate to positional accuracy.

The interpolation order also affects the performance of our variants. The \texttt{-r} configuration performs best on most rotational and update-regime metrics, whereas \texttt{-d} achieves the lowest MPJPE. This supports the expected influence of the hierarchical fusion order, although the differences between variants remain relatively small on several pose-level criteria.

For sequence- and event-based metrics, \texttt{RPM} obtains the best score on three of the four criteria, while \texttt{REACT-learned-2S4-r} achieves the lowest
PJ$_{T\to S}$ and ranks second on the remaining three. Conversely, the GRU baselines and distilled students exhibit substantially larger PJ and AUJ values, indicating less stable motion during changes between tracking and synthesis regimes.

\begin{table}[t]
\caption{Motion-completion performance of \emph{REACT} variants,
distilled students, and causal baselines. Lower is better for all
reported metrics.}
\centering
\small
\setlength{\tabcolsep}{1.5pt}
\renewcommand{\arraystretch}{0.6}
\begin{subtable}{\linewidth}
\caption{Pose-level metrics}
\label{tab:fra_met}
\centering
\begin{tabular}{lcccc}
\toprule
Model & \makecell[c]{MPJPE \\ \text{[cm]}}  & \makecell[c]{P-MPJPE \\ \text{[cm]}}  & \makecell[c]{MPJLAE \\ \text{[deg]}}  & \makecell[c]{MPJAE \\ \text{[deg]}}  \\
\midrule
\midrule

RPM  & $11.1 \pm 6.9$ & $8.7 \pm 4.1$ & $24.2 \pm 4.7$ & $30.8 \pm 18.6$  \\

\midrule
GRU-ctx & $8.8 \pm 6.4$ & $6.6 \pm 4.1$ & $22.7 \pm 5.5$ & $27.0 \pm 15.9$ \\

GRU-ctx-large & $11.0 \pm 6.7$ & $8.4 \pm 4.0$ & $20.6 \pm 5.1$ & $28.3 \pm 19.7$ \\

\midrule

GRU-ctx-student & $10.6 \pm 7.0$ & $8.0 \pm 4.1$ & $14.4 \pm 5.6$ & $24.8 \pm 20.4$\\

VQVAE-ctx-student & $10.9 \pm 7.4$ & $8.5 \pm 4.6$ & $13.9 \pm 6.1$ & $23.9 \pm 21.3$ \\

\midrule
REACT-learned-2S4-r& $\msecond{7.8 \pm 6.3}$ & $\mbest{5.9 \pm 4.2}$ & $\mbest{12.2 \pm 6.6}$ & $\mbest{19.6 \pm 15.4}$  \\
REACT-learned-2S4-p & $7.9 \pm 6.5$ & $6.0 \pm 4.2$ & $14.9 \pm 6.7$ & $20.6 \pm 17.1$  \\
REACT-learned-2S4-d & $\mbest{7.7 \pm 6.4}$ & $\msecond{5.9 \pm 4.2}$ & $\msecond{13.8 \pm 6.7}$ & $\msecond{20.1 \pm 15.5}$ \\

\bottomrule
\end{tabular}

\end{subtable}

\begin{subtable}{\linewidth}

\caption{Sequence- and event-based metrics}
\label{tab:seq_met}
\centering
\begin{tabular}{lcccc}
\toprule
Model & \makecell[c]{PJ$_{T\to S}$} & 
\makecell[c]{PJ$_{S\to T}$} &
\makecell[c]{AUJ$_{T\to S}$} &
\makecell[c]{AUJ$_{S\to T}$} \\
\midrule
\midrule

RPM  & $\msecond{1.5 \pm 5.6}$ & $\mbest{1.8 \pm 7.6}$ & $\mbest{25.4 \pm 13.1}$ & $\mbest{26.4 \pm 22.3}$\\

\midrule

GRU-ctx & $5.2 \pm 3.6$ & $8.7 \pm 6.3$ & $105.3 \pm 47.3$ & $105.3 \pm 46.1$ \\

GRU-ctx-large & $6.4 \pm 3.9$ & $10.0 \pm 6.8$ & $134.4 \pm 70.6$ & $135.3 \pm 63.9$ \\

\midrule

GRU-ctx-student & $5.6 \pm 4.7$ & $11.6 \pm 7.8$ & $104.7 \pm 48.2$ & $127.1 \pm 53.8$\\

VQVAE-ctx-student & $6.6 \pm 5.3$ & $14.2 \pm 9.6$ & $128.4 \pm 57.7$ & $148.8 \pm 63.6$ \\

\midrule
REACT-learned-2S4-r& $\mbest{1.4 \pm 1.9}$ & $\msecond{2.0 \pm 3.4}$ & $\msecond{25.6 \pm 5.7}$ & $\msecond{26.8 \pm 10.3}$\\
REACT-learned-2S4-p & $2.1 \pm 2.8$ & $3.8 \pm 5.0$ & $29.9 \pm 10.0$ & $33.3 \pm 18.7$\\
REACT-learned-2S4-d & $1.7 \pm 1.5$ & $2.7 \pm 2.7$ & $27.4 \pm 6.7$ & $29.4 \pm 9.1$ \\

\bottomrule
\end{tabular}
\end{subtable}
\end{table}

\begin{figure}[ht]
    \centering
    \includegraphics[width=1\linewidth]{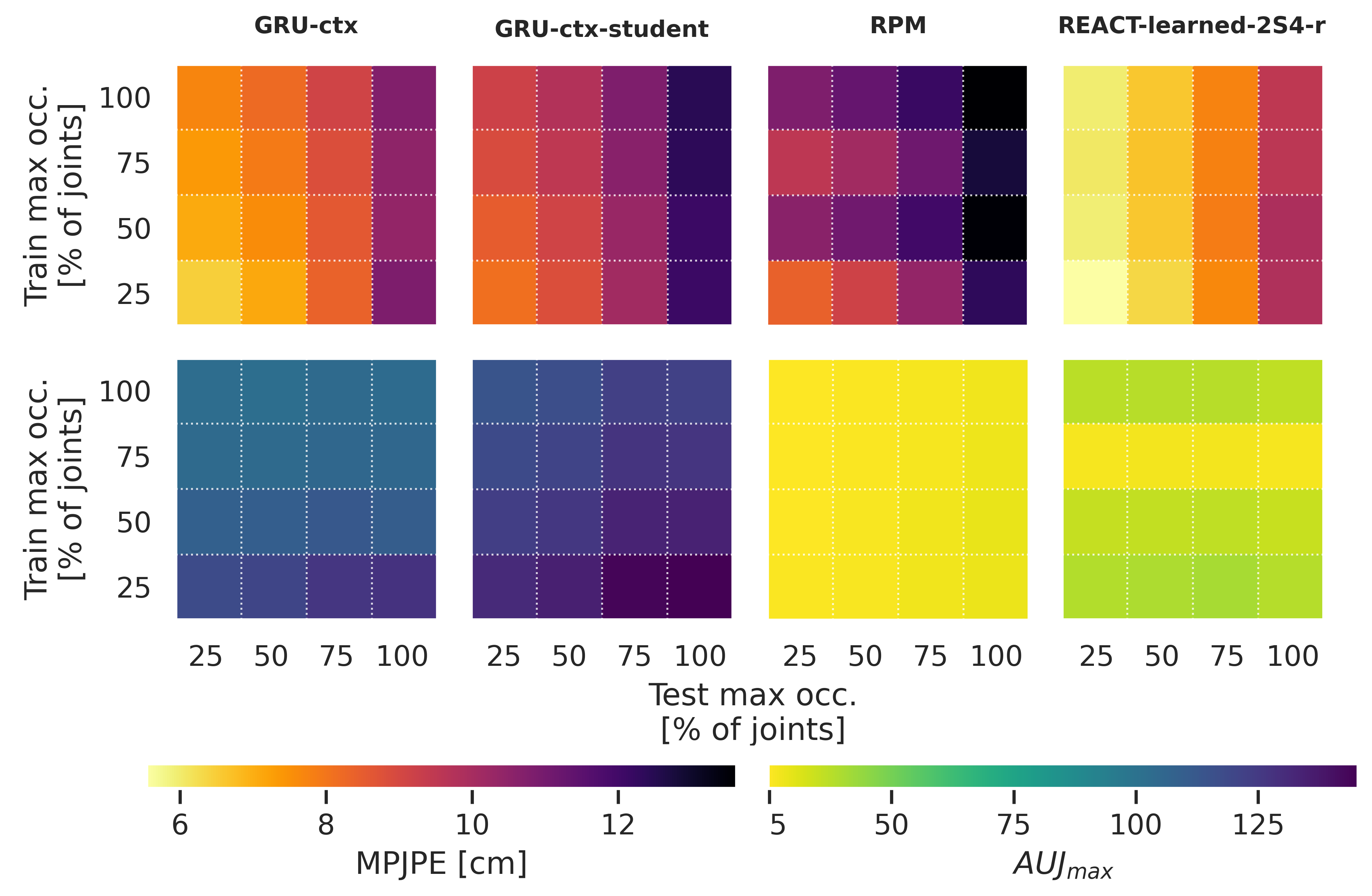}
    \caption{Robustness to masking severity. Rows correspond to $M_{\max}^{\text{train}} \in \{25,50,75,100\}\%$ and columns to $M_{\max}^{\text{test}} \in \{25,50,75,100\}\%$. Each model cell reports the mean MPJPE (top) and $AUJ_{\max}=\max(AUJ_{T\to S},AUJ_{S\to T})$ (bottom), with lower values indicating better performance for both metrics. Increasing $M_{\max}$ allows larger joint masks, up to fully
    autoregressive segments.}
    \label{fig:maskrobustnessshort}
\end{figure}

\subpara{Mask severity robustness}
This evaluation quantifies robustness to mismatches between the maximum masking severity used during training and testing. For each architecture, we train four checkpoints with $M_{\max}^{\text{train}} \in \{25,50,75,100\}\%$ and evaluate them under four test regimes with $M_{\max}^{\text{test}} \in \{25,50,75,100\}\%$. We focus on two complementary metrics: MPJPE for pose reconstruction and $AUJ_{\max}=\max(AUJ_{S\to T},AUJ_{T\to S})$ for motion stability during update regimes. Although these two metrics do not capture all aspects of completion quality, they provide a compact view of how reconstruction accuracy and update dynamics evolve with mask severity (\cf~Figure~\ref{fig:maskrobustnessshort}).

Across all models, MPJPE is primarily affected by $M_{\max}^{\mathrm{test}}$, confirming that increased test-time masking severity makes reconstruction more difficult. For most architectures, MPJPE also tends to increase with $M_{\max}^{\mathrm{train}}$, although this effect is comparatively smaller.

The GRU-based models exhibit a clear trade-off. Training with a lower $M_{\max}^{\mathrm{train}}$ yields lower MPJPE under mild test-time masking, but generally results in larger $AUJ_{\max}$. Conversely, increasing the training masking severity improves update smoothness at the expense of per-frame reconstruction accuracy. The distilled GRU student exhibits a similar pattern, suggesting that this sensitivity depends at least partly on the underlying architecture.

Consistent with the previous evaluation, \texttt{RPM} achieves the lowest $AUJ_{\max}$ at the cost of higher MPJPE. By contrast, the Ref-last (\texttt{-r}) maintains relatively low values for both metrics across the evaluated settings. In particular, for a fixed $M_{\max}^{\mathrm{test}}$, its MPJPE varies comparatively little as $M_{\max}^{\mathrm{train}}$ changes. This suggests reduced sensitivity to the training masking severity. Additional results and analysis are reported in the supplementary material.

\begin{figure}[ht!]
    \centering
    \includegraphics[width=1\linewidth]{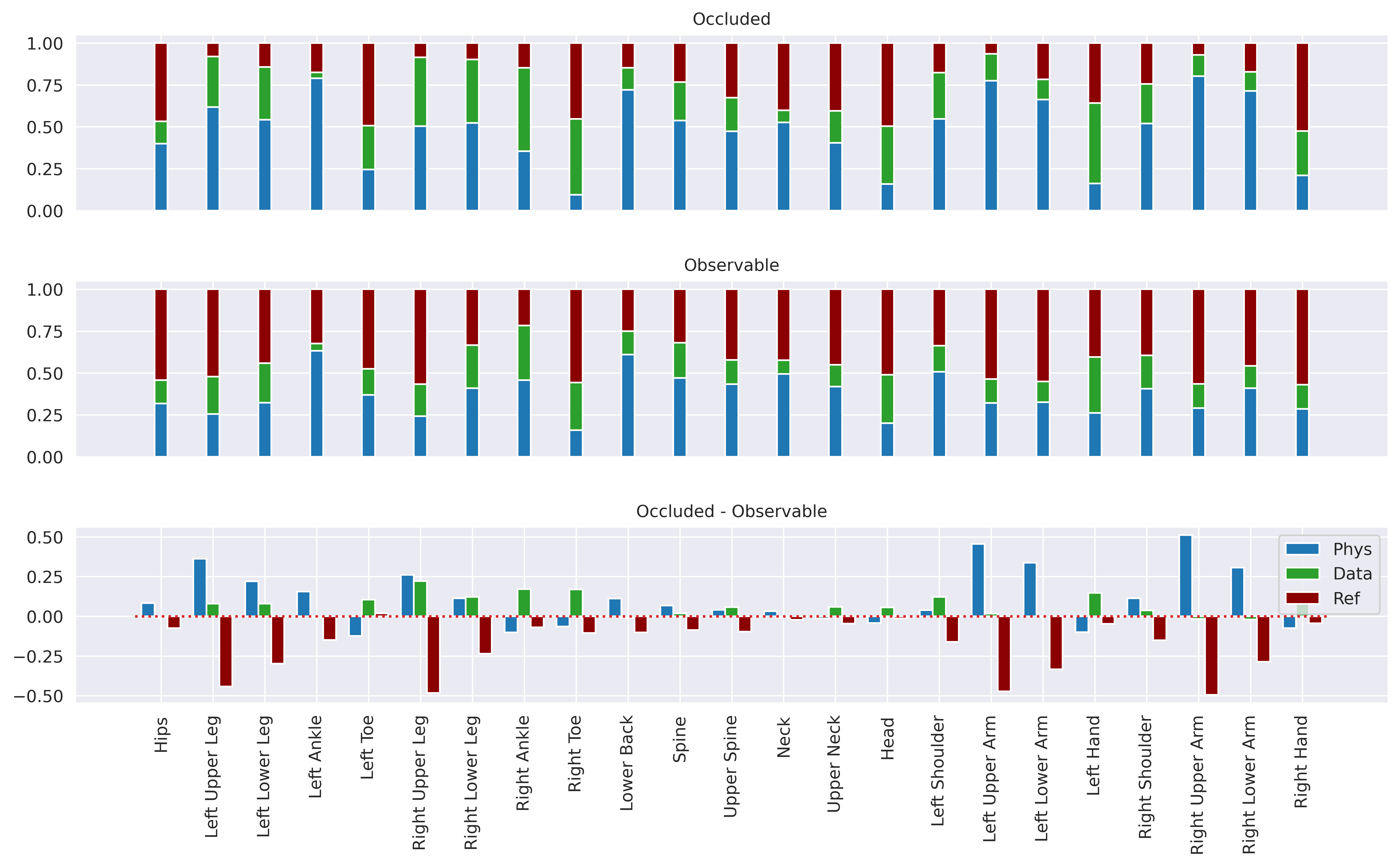}
    \caption{Per-joint average fusion weights for the Ref-last (\texttt{-r}) variant under ``Occluded'' (top) and ``Observable'' (middle) conditions. The difference map ``Occluded--Observable'' is shown at the bottom, where positive values indicate an increased contribution under occlusion.}
    \label{fig:mix_occobs}
\end{figure}

\para{REACT fusion weights}
Finally, we briefly examine the internal behaviour of \emph{REACT} through its fusion weights. Figure~\ref{fig:mix_occobs} reports the per-joint average contribution of each completion candidate conditioned on the ground-truth occlusion state. Under occlusion, the contribution of \emph{Ref} consistently decreases, while the predictive candidates, primarily \emph{Phys}, receive larger weights. This behavioural shift is consistent with the intended reliability-aware coordination of the three motion sources. A more detailed analysis of fusion dynamics and reliability estimation is provided in the supplementary material.

\section{Discussion}
\label{sec:Discussion}
Our results highlight distinct behaviours among the evaluated model families. GRU-based models achieve competitive per-frame reconstruction but exhibit substantially less stable motion during update regimes. Conversely, \texttt{RPM} converges toward a conservative, low-pass-filter-like behaviour, producing smooth motion at the cost of responsiveness and reconstruction accuracy. \emph{REACT} provides an intermediate trade-off, with comparatively low reconstruction errors and update-smoothness metrics close to \texttt{RPM}, while exposing interpretable per-joint signals and reliability-dependent contributions from its three motion sources. This decomposition nevertheless shows limitations. Even with injected priors, the \emph{Data} branch rarely dominates during occlusions, while \emph{Phys} remains the preferred predictive candidate in many cases. This may result from the greater ambiguity of longer-horizon prediction and the difficulty of jointly optimizing competing candidates. A more explicitly generative objective, conditioned on reliable observations and coupled with a stronger lightweight generative architecture, could better differentiate its role. 

Although motivated by interactive and ultimately real-time applications, our evaluation relies on several assumptions.

First, we use synthetic corruptions on a consistent skeletal representation. Our augmentation includes rotation-aware noise and temporally persistent joint- and body-part masks, but does not reproduce persistent drift, acquisition-frequency variations, tracker synchronization errors, or upstream pose-estimation failures. Evaluation on real tracking failures is therefore required to establish practical robustness.

Second, all experiments use a fixed and normalized skeletal topology and do not account for morphological variation or intra- and cross-structural retargeting. Extending the method to heterogeneous skeletons would broaden its applicability.

Third, our variants operate at approximately 45~fps on an RTX 3090 (\cf~Table~\ref{tab:complexity}). This supports interactive use under favourable conditions but leaves limited computational margin and does not meet the 60--90~Hz rates commonly targeted by immersive systems. Reduced-precision inference, compilation or fusion of custom operations, and simplification of the rotation-processing pipeline remain unexplored optimizations. For instance, rotation conversions and related geometric operations alone introduce approximately $3.86$~ms of overhead ($17.6\%$ of \texttt{REACT-learned-2S4} latency). More broadly, identifying which structural components are responsible for the observed performance would provide a basis for designing a lighter model with a larger latency margin.

Fourth, our evaluation remains primarily reconstruction-oriented, despite the partly generative role assigned to the system during extended occlusions. Standard reconstruction metrics do not necessarily correlate with human perception~\citep{rekik_qa_2025} and may therefore fail to capture perceptually implausible motion. Moreover, because the system processes observations even when they are reliable, it can introduce model-induced distortions during the transmission regime. Future evaluations should therefore complement reconstruction metrics with perceptual and biomechanical criteria, and separately assess the different operating regimes: direct observation, short- and long-horizon completion, and the updates between them. 

\section{Conclusion}
\label{sec:Conclusion}
In this work, we present a new formulation and corresponding architecture for causal and interactive motion completion under unknown spatiotemporal sparsity and noise. Our method, called \emph{REACT}, can be understood as the reliability-aware coordination of three completion candidates: a directly transmitted observation, a short-term prediction, and a generated longer-horizon estimate. This core formulation builds on several primitives from the literature, adapted to our setting, including an input-filtering module and a PCAF-inspired, geometry-aware correction applied to a rolling buffer.
Our evaluations suggest that \emph{REACT} provides a favourable trade-off between reconstruction accuracy and temporal stability, while exhibiting clear behavioural shifts depending on the occlusion state and exposing interpretable internal signals.


\end{document}


\title{REACT: Towards Real-Time Completion of Skeletal Motion Signals \\-- Supplementary Material --}

\maketitle

\section{Organization}
In this supplementary material, we present additional experiments that are relevant to both the task and the method presented in the main paper.

\begin{landscape}
\begin{figure}[p]
    \centering
    \includegraphics[width=\linewidth]{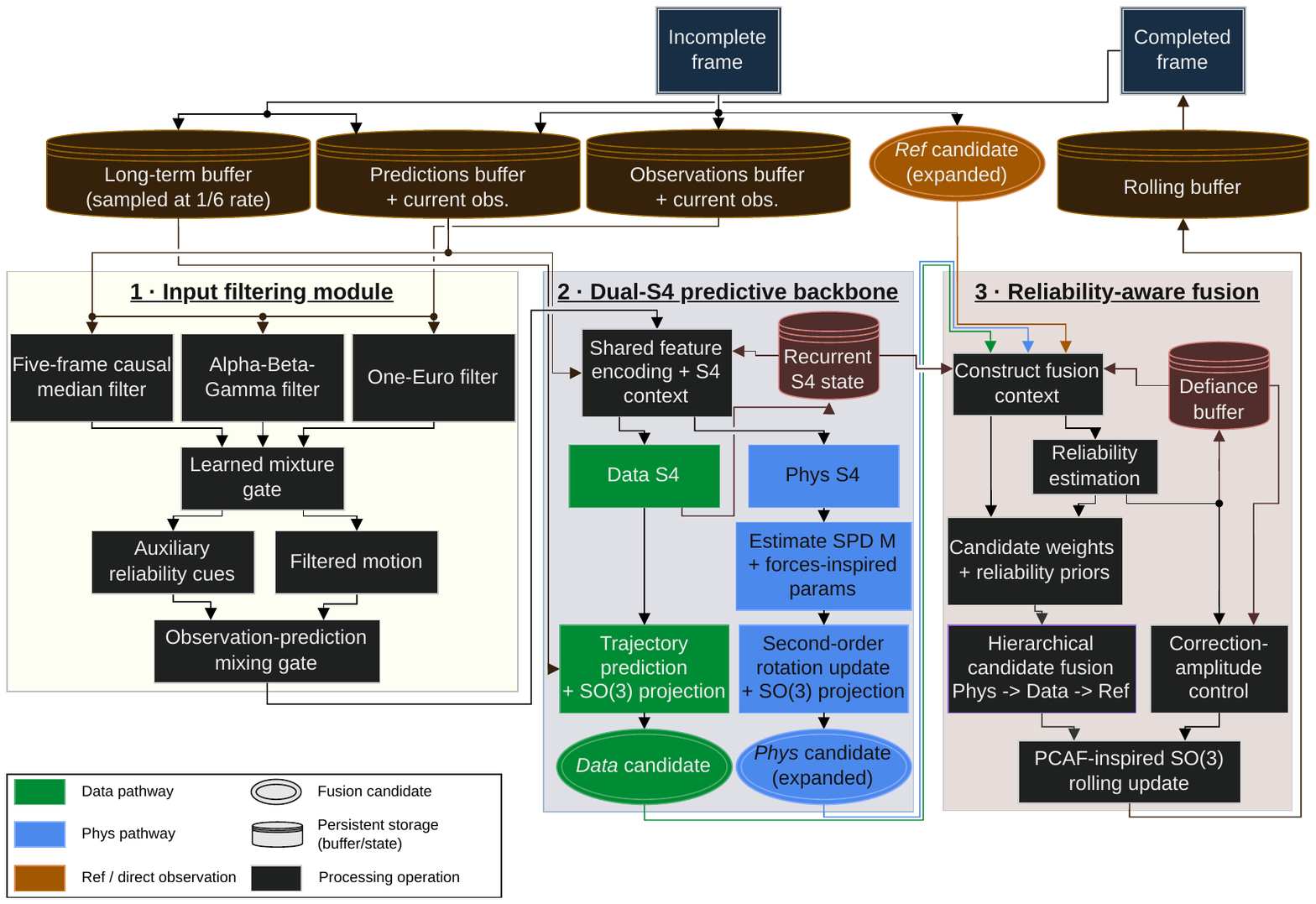}
    \caption{Overview of the REACT architecture and, in particular, of the \texttt{2S4-r} variant. We represent the three main modules, the different states and buffers and the general behaviour of the method. Colours identify the \emph{Data}, \emph{Phys}, and \emph{Ref} pathways,
    while node shapes distinguish fusion candidates, persistent states or
    buffers, and processing operations.}
    \label{fig:react_overview}
\end{figure}
\end{landscape}

First, we provide additional architectural details (Section~\ref{sec:supp_partitionedS4}) that clarify several design choices and limitations. This in-depth description leads to the presentation of several design choices and ablations that rely on lower-level implementation details, such as the use of a single partitioned S4 layer as backbone (\cf~Figure~\ref{fig:dual_head}). We also provide a complete overview of the method in Figure~\ref{fig:react_overview}.

Second, as advertised in the main paper, we formally define our metrics and how we prioritize them during our experiments in Section~\ref{subsec:supp_metrics}.

Third, we introduce a more exhaustive evaluation in Section~\ref{sec:supp_experiments}. In particular, we provide comparisons of the newly defined variants against the baselines that are introduced in the main paper. To better situate the performance of the variants presented in the main paper in this new framing, we also include their results.
We also design another set of baselines with similar architectures in a context without the observations buffer. The results of this task-level ablation, although partial, may yield relevant insights for possible future works on the completion problem.

Fourth, to better characterize our method, we report its behaviour under occlusion on the entire test set (Section~\ref{subsec:supp_internal}). For instance, we present an analysis of the fusion-weights dynamics and, ultimately, visualize the performance of the reliability estimation mechanism.

\section{Additional architecture details}
\label{sec:supp_partitionedS4}
In this section, we present several low-level details that may be of interest to the reader to reproduce the method.

Most linear transformations within the MLPs are implemented as low-rank layers, factorizing the usual weight matrix into the product of two smaller matrices. This reduces the parameter count while preserving the expressive capacity of the surrounding non-linear mapping. In our experiments, this factorization did not lead to a clear degradation in performance compared with standard fully connected layers and, in several cases, was associated with more stable optimization during training.

Beyond the current input, our method maintains several persistent states and history buffers across inference steps. For clarity, Table~\ref{tab:buffers} summarizes these persistent variables and their respective roles.

\begin{table}[t]
\caption{Persistent states and history buffers maintained by our system. Sequence lengths $T=31$ and $T_{r}=10$, number of joints $J=23$, rotation representations $F=6$ (6D) and $F_{r}=9$. The latter is a flattened rotation matrix, which leads to both an explicit and unambiguous rolling buffer and an easier rotation composition when necessary. Backbone features $H=256$, backbone modes $N=32$.}
\label{tab:buffers}

\centering
\small
\setlength{\tabcolsep}{5pt}
\begin{tabular}{lcccc}
\toprule
Buffer & Size & Module & Objective & Details \\
\midrule
\midrule
Predictions   & $\mathbb{R}^{T \times J \times F}$ & Input & Autoregressive context & \makecell[c]{Previous outputs\\and current input} \\
\midrule
Observations &  $\mathbb{R}^{T \times J \times F}$ & Input & Observation history & \makecell[c]{Stacks raw\\measurements}  \\
\midrule
\makecell[l] {Long-Term} &  $\mathbb{R}^{T \times J \times F}$ & S4 Fusion &  \makecell[c] {Long-term context} &  \makecell[c] {Samples predictions\\at lower frequency }\\
\midrule
S4 state & $\mathbb{C}^{H \times N}$ & S4 Fusion & \makecell[c] {Compact context} & \makecell[c] {Backbone output,\\complex-valued \\ (diagonalization\\in complex plane)} \\
\midrule
Defiance & $\mathbb{R}^{J}$ & Fusion & \makecell[c] {Joints trust \\ history/hysteresis} &  \makecell[c] {Up if low joint confidence,\\else down}\\
\midrule
Rolling &  $\mathbb{R}^{T_r \times J \times F_r}$ & Fusion & \makecell[c] {Temporal\\correction} & PCAF-like update\\

\bottomrule
\end{tabular}
\end{table}

Next, we present an architecture configuration complementary to the one evaluated in the main material and we report its performances in Section~\ref{sec:supp_experiments}.

\subsection{Partitioned S4}
\label{subsec:supp_S4}
In the main paper, we introduce a backbone using two parallel S4 layers. Both layers independently process a common input embedding before their branch-specific heads. Alternatively, the structure of a single S4 layer naturally allows its modes to be shared or partitioned between multiple output heads, yielding a more strongly shared latent representation.

\begin{figure}[t!]
    \centering
    \includegraphics[width=0.95\linewidth]{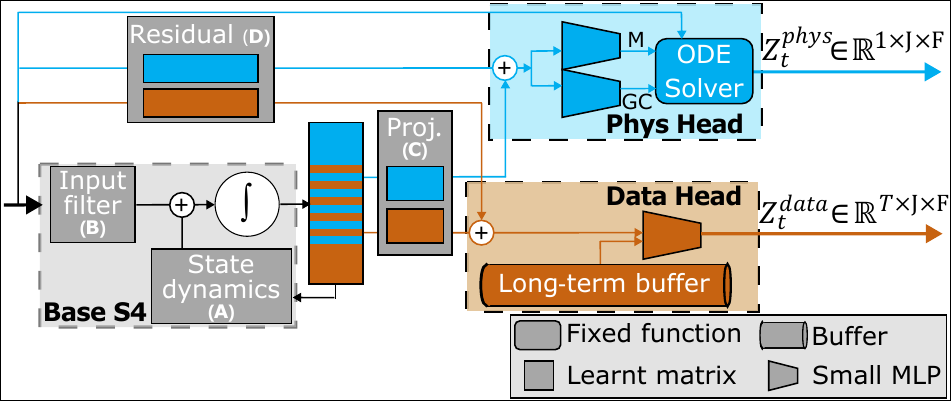}
    \caption{Base S4 and constrained dual-head outputs. The model processes a sequence of features using shared S4 dynamics and constructs two convolution kernels through distinct output mappings. The resulting features are then processed by two branch-specific heads and objectives. The first, depicted in light blue, uses an ODE-inspired
    parameterization~\citep{Zhang_2024_WACV} for short-term prediction. The second, depicted in orange, uses a long-horizon context to regress a sequence of future predictions.}
    \label{fig:dual_head}
\end{figure}

The standard continuous-time state-space formulation underlying S4 is
given by
\begin{equation}
    \left \{
    \begin{array}{cc}
        \dot x = Ax + Bu \\
         y = Cx + Du
    \end{array},
    \right .
\end{equation}
where $x$ is the state vector and $u$ the input signal. We retain the same state dynamics but introduce two distinct output mappings:
\begin{equation}
    \left \{
    \begin{array}{r c l}
        \dot{x} &=& Ax + Bu, \\
        y_{\mathrm{phys}} &=& C_{\mathrm{phys}}x + D_{\mathrm{phys}}u \\
        y_{\mathrm{data}} &=& C_{\mathrm{data}}x + D_{\mathrm{data}}u
    \end{array}  
    \right .
\end{equation}

Because the underlying dynamics are shared, unconstrained optimization could allocate most of the useful representational capacity to one branch at the expense of the other. To encourage both branches to retain dedicated capacity while exploiting the spectral structure of the state-space model, we partition the S4 modes across the two output heads. Let $N$ denote the number of modes in the state matrix $A$. Both heads share $N/2$ modes, while each additionally receives exclusive access to $N/4$ ``private'' modes. In practice, we implement this by zero-padding the corresponding mode components of the output mappings $C_{\mathrm{phys}}$ and $C_{\mathrm{data}}$. 

At initialization, standard S4 modes can be viewed as approximately organized according to their characteristic timescales. We exploit this structure to introduce a loose initial spectral bias between the two branches. Importantly, no such separation is enforced after initialization, and the initial specialization may therefore be attenuated or disappear during training. Conceptually, the two heads share 50\% of the latent modes, while the remaining two 25\% subsets provide branch-specific capacity (\cf~Figure~\ref{fig:dual_head}). 

The remainder of the layer follows the standard S4 computation, except that two convolution kernels are produced instead of one, while retaining the same underlying mathematical optimizations.

Finally, we use S4 in convolutional rather than recurrent form because we retain an explicit short-term sequence context. This increases inference cost, but the backbone also provides a compact S4 state that is subsequently used as an auxiliary context signal by the fusion module.

\subsection{Static gate}
\label{subsec:supp_static_gate}
In the main paper, we propose to merge the input filters results with a \texttt{learned} gate. In our experiments, we also designed a \texttt{static} one, which is based on human-readable and interpretable signal-derived cues.

Broadly, our \texttt{static} gate regresses the filter weights based on the energy of the error signal and the input dynamics. More specifically, for each filter we first compute its residual $e_{filter} = \vert y_{filter} - x\vert$ and robustly scale it (based on its median) to define some logits $z_{coarse}=-\text{scaling}(e_{filter})$. At this point, the gate is naive and prioritizes the filter with a small error, which may impair the filters efficiency. As such, in the same spirit as the One-Euro filter, we leverage the motion velocity to adjust the priors. We smooth the absolute value of the velocity using an Exponential Moving Average (EMA) and normalize it, which results in a signal $E_n$. When this signal is high, we increase the weight of the median filter since it may reflect the velocity spikes of an artifact. In contrast, when $E_n$ is low, we give more weight to the other, more flexible filters. We add these second priors (denoted $b_{filt}$) and a dynamical term generated by a small trained convolution layer $g_\theta$ to the previous logits and obtain $z_{gate} = z_{coarse}+b_{filt}+g_\theta(E_n)$, which is then fed to a softmax to compute normalized probabilities.
Note that here, ``static'' refers to the explicitly designed structure and priors of the gate and a small learned dynamic correction is nevertheless retained.

\section{Experiments}
\label{sec:supp_experiments}
In this section, we present additional experiments and ablations.

\subsection{Evaluated models}
\label{subsubsec:supp_models}

\subsubsection{Variants}
We define here all the variants defined by our newly presented design choices. We already introduced \texttt{REACT-learned} variants, and we thus define \texttt{REACT-static} ones, which use the \texttt{static} gate.

In addition to the \texttt{-2S4} models, we want to observe the effect of partitioning a S4 layer between both heads, as described in Section~\ref{subsec:supp_S4}. In addition, we can also observe the effect of loosely biasing the branches towards distinct spectral modes in the initial steps of the training. Thus, \texttt{-lfd\_hfp} indicates an initial bias towards low frequencies (lf) for the \emph{Data} (d) branch and high frequencies (hf) for the \emph{Phys} (p) branch. By contrast, \texttt{-lfp\_hfd} emphasizes an initial bias towards low frequencies for the \emph{Phys} branch and high frequencies for the \emph{Data} branch.

\subsubsection{Observations buffer}
To measure the effect of the observations buffer on the completion task, we articulate the training / inference of our problem under two distinct configurations. Each defines its own inputs and, as such, a set of models that can be compared fairly.

\subsubsection{Configuration 1: Two-buffer input}
That is precisely the configuration used in the main paper. Here, models receive both the prediction buffer and the raw-input buffer (\ie, noisy and possibly missing measurements) and we describe them slightly more.

This configuration is used to train three GRU baselines, two distilled students and a rolling prediction model. We start by describing the GRUs.

(i) \texttt{GRU-ctx} is a 3-layer GRU with hidden size $FJ$, preceded by a linear projection that maps the concatenated buffers to the GRU input size,

(ii) \texttt{GRU-ctx-large} follows the same structure but uses a 5-layer GRU with hidden size 512,

(iii) \texttt{GRU-ctx-large-noproj} removes the input projection and feeds the (higher-dimensional) concatenated buffers directly to a 5-layer GRU with hidden size 512.

We report two distilled students with the same two-buffer input format and input linear projection: \texttt{VQVAE-ctx-student} (VQVAE backbone) and \texttt{GRU-ctx-student} (same backbone as \texttt{GRU-ctx}).

\subsubsection{Configuration 2: One-buffer input} 
In this configuration, models receive only the prediction buffer and the current incomplete frame. We evaluate models, without input projection, including:

(i) \texttt{GRU}, a 3-layer GRU with hidden dimension $FJ$,

(ii) \texttt{GRU-large}, a 5-layer GRU with hidden dimension 512,

(iii) two students \texttt{GRU-student} and \texttt{VQVAE-student}.

This configuration is closer to a pure prediction setup and we expect it to reduce processing at runtime as well as memory footprint.

Strictly speaking the two configurations differ in formulation (\ie, inputs dynamical filtering \vs prediction-centric completion). As such, comparing them may also provide practical insights into the trade-off between robustness and simplicity in our completion task.

\subsection{Metrics}
\label{subsec:supp_metrics}
In this section, we exhaustively present and define the metrics used to quantify the performance of our trained models \wrt the completion task challenges. We hierarchize our criteria into two categories: the primary metrics (our main objectives) and more complementary scores, highly dependent on context, to provide a deeper analysis.

\subsubsection{Primary metrics}
\label{subsubsec:prim_met}
The metrics disclosed in this section are standard in the motion reconstruction literature. When applicable, they are scaled to obtain figures making sense in the International System of Units, using our skeleton's normalized offsets. We robustly approximate a factor by using the distance between the head and toes of a standing skeleton over several frames and scaling it to reach $170\,\text{cm}$.

Among our reported metrics, we include scores on the position, such as the MPJPE (in world coordinates) defined as follows:
\begin{equation}
    \text{MPJPE} = \frac{1}{J}\sum_{j=0}^{J-1}\lVert p_j^t - \hat{p}_j^t\rVert_2
\end{equation}, with $\hat{p}_j^t$ and $p_j^t$ the predicted and ground-truth joint positions. We also use the Procrustes-aligned MPJPE (P-MPJPE), which removes the optimal global similarity transformation before
computing the positional error via a Singular Value Decomposition (SVD).

We also address joint rotation errors through the geodesic distance: 
\begin{equation}
D_{\mathrm{geo}}(R,\hat R)
=
\arccos\!\left(
    \frac{\operatorname{tr}(R\hat R^\top)-1}{2}
\right)
\end{equation}
between $\hat{R}$ and $R$, respectively the predicted and ground-truth rotations, applied on both local (MPJLAE) and global rotations (MPJAE). We emphasize their interest here since, although implicitly encoding rotational information, position-only metrics cannot reliably assess errors over all rotational Degrees of Freedom (DoFs). Indeed, such metrics are computed by collapsing sets of joint rotations into identical positional configurations and thus cannot disambiguate errors along axes that do not affect them. For instance, a rotation around a bone axis does not change the child joint position in space, even though creating artifacts during automatic skinning (\eg, Linear Blend Skinning). As another example, position-only metrics often fail to reliably measure errors on distal joints \citep{kucherenko_mocap_2025}, which may ultimately lead to models producing extreme failure scenarios (\eg, $> 90\degree$ wrist bending).

To further measure the reliability of our system as a filter in an interactive setting, we use the metrics introduced by Barquero~\etal~\citep{barquero_auj_2024} to measure discrepancies during the update regime: the Peak Jerk (PJ) and the Area Under the Jerk (AUJ) such that:
\begin{equation}
    \text{PJ} = \max_{\substack{1 \leq i \leq K \\ 1 \leq \tau \leq L_{ur}}} \lVert j_i(\tau) \rVert_{1},
\end{equation}
\begin{equation}
    \text{AUJ} = \sum_{\tau=1}^{L_{ur}} \max_{1 \leq i \leq K} \lVert j_i(\tau) - j_{\text{avg}} \rVert_{1},
\end{equation} with $L_{ur}$ the length of the measured update regime ($\approx 1\,\text{s}$), $K$ the number of joints, $j_i(\tau)$ the jerk of the $i^{th}$ joint at time $\tau$ and $j_{avg}$ the average joint-wise maximum jerk across the dataset ($0.3844$ in our case).

Motion dynamics discrepancies are penalized by MPJPVE and MPJAVE. They represent the Mean Per Joints Velocity Error (even though sometimes denoted MPJVE in the literature, we call it MPJPVE to prevent any confusion with the next metric) and the Angular Velocity (or MPJAVE for Mean Per Joints Angular Velocity Error). They are defined as follow: 
\begin{equation}
    \text{MPJPVE} =
    \frac{1}{J}
    \sum_{j=0}^{J-1}
    \lVert v_j^t - \hat{v}_j^t \rVert_2,
\end{equation}
\begin{equation}
    \text{MPJAVE} =
\frac{1}{\Delta t\cdot J}
\sum_{j=0}^{J-1}
\left|
D_{\mathrm{geo}}(R_j^t,R_j^{t+1})
-
D_{\mathrm{geo}}(\hat{R}_j^t,\hat{R}_j^{t+1})
\right|,
\end{equation}
with $v_j^t = \frac{p_j^{t+1}-p_j^t}{\Delta t}$, $D_{geo}$ as defined previously and $\Delta t = \frac{1}{60}$ in our case.
Time derivative metrics tend to be unstable, with high inter-animation variance and may become arbitrarily large if a model sometimes produces velocity vectors of opposite direction \wrt ground-truth. For these reasons, we consider them of intermediary importance between primary and complementary metrics.

\subsubsection{Complementary metrics}
\label{subsubsec:comp_met}
In addition to the per-frame (and per-event, for the PJ/AUJ) metrics, we introduce per-sequence scores. They aim to measure a similarity between the model's output structure and the ground-truth's. Specifically, we try to estimate whether the trained completion models preserve the spatiotemporal organization of the ground-truth motion by analysing their residual. For instance, an over-smoothing model may achieve low MPJPE and AUJ, while failing to reproduce high-frequency components. An extreme example of such a phenomenon would be a model outputting a static, average pose for most joints, while providing a robust rotation of proximal joints (\eg, hips and trunk).

\para{Temporal Power Spectral Entropy (TPSE)}
This metric corresponds to the entropy of the power spectral density distribution, and is computed on the temporal axis. Let $p,\hat{p} \in \mathbb{R}^{T_{psd} \times J \times 3}$ be the ground-truth and predicted joint positions respectively, and define the residual (or error signal) $r=p-\hat{p}$. We flatten it over joints to obtain $r_f \in \mathbb{R}^{T_{psd} \times 3J}$. We compute its power spectral density $\text{PSD}(r_f)$ (\eg, via Welch's method) and define the normalized spectral distribution $p_i=\frac{\text{PSD}(r_f)_i}{\sum_j\text{PSD}(r_f)_j}$. The temporal spectral entropy is then:
\begin{equation}
    \text{TPSE}=\frac{-\sum_i p_i \log_2(p_i)}{H_{max}},\ H_{max} = \log_2(K),
\end{equation} where $K$ is the number of frequency bins so that $H_{max}$ is the maximum entropy of our spectrum. In other words, the TPSE aims to assess if the error signal still includes a temporal structure, \ie, a non-uniform distribution of the signal power between low- and high-frequencies.

\para{Residual Compression Size (RCS)}
This metric aims to measure a complementary notion of residual complexity. As its name suggests, we use existing algorithms to compress the quantized residual and compare the resulting size, \ie, let $r_q$ be the quantized version of $r_f$ as defined previously. Then $\text{RCS}=C(r_q)$ with \eg, $C=\text{gzip}$. Such a compression process is already designed to capture relevant patterns, and we leverage it to loosely indicate the residual complexity (\eg, a smooth and low-variance residual would have a lower compressed complexity than a highly irregular and noisy one). Note that this criterion is vulnerable to the inherent complexity of the motion (\eg, a T-Pose would be much easier to compress than dancing), which implies high inter-motion and thus high intra-model variance.

\para{Limitations} In practice, we only compute both criteria described previously on short sequences ($\approx15-20$ frames). This choice is motivated by the presence of real capturing noise in the ground-truth of the dataset, which could have invalidated our results. As such, we choose to compute the signal-complexity metrics on short sliding windows, which means most of the low-frequency content is missing (thereby mitigating the compressing capacities of gzip). Furthermore, absence of structure in the residual does not directly imply a good model, nor does structure necessarily tie to failure, in other words TPSE and RCS do not have a universally preferable direction, which motivates their use as complementary metrics.

\subsection{Quantitative results}
\subsubsection{Complexity}
We summarize the results of our experimented models' complexity evaluation in Table~\ref{tab:annex_complexity}. 
Despite the higher inference cost, our variants exceed the 30~fps threshold, which may be suitable for some interactive applications. However, they are currently not fast enough to support 60 or 90 Hz in particular under increased hardware constraints (\eg, smaller GPUs). We observe that the one- and two-S4 variants exhibit similar runtime costs, suggesting that a substantial fraction of REACT's latency originates outside the backbone itself. In short, a reasonable hypothesis is that the higher inference latency is caused by our non-fused custom operations, which lead to multiple CUDA kernels calls. %
As expected, we observe that the models in the second configuration are consistently faster than their equivalent in the first configuration (\eg, \texttt{GRU} \vs \texttt{GRU-ctx}), as they do not need to maintain an additional input buffer.

\begin{table}[ht!]
\caption{Complexity and runtime. Hardware: RTX 3090 (24 GB), batch size 1, sequence length 31, joints dimension 23, precision FP32. Environment: PyTorch 2.3.1, CUDA 12.1, cuDNN 8.9.2. Metrics are aggregated over 500 iterations following 100 steps of GPU warm-up. We further report the 50th and 95th percentiles (respectively P50 and P95).
}
\label{tab:annex_complexity}

\centering
\small
\setlength{\tabcolsep}{3pt}
\begin{tabular}{lccc}
\toprule
Model & \makecell[c]{Params [M]} & 
\makecell[c]{Latency [ms] \\ P50 | P95}  & 
\makecell[c]{VRAM [MB]} \\
\midrule
\midrule
RPM   & 6.75 & 4.03 | 4.30 & 98 \\
\midrule
GRU-ctx & 0.38 & 0.57 | 0.64  & 24\\
GRU-ctx-large & 8.09 & 1.41 | 1.60 & 148\\
GRU-ctx-large-noproj & 7.59 & 1.42 | 1.65 & 142\\
\midrule
VQVAE-ctx-student & 0.38 & 1.47 | 1.78 & 24 \\
GRU-ctx-student & 0.38 & 0.62 | 0.66 & 22 \\
\midrule
REACT-static & 1.05 & 22.55 | 24.12 & 242 \\
REACT-learned & 1.05 & 21.65 | 22.98 & 222 \\
REACT-learned-2S4 & 1.10 & 21.94 | 23.71 & 222 \\
\midrule
\midrule
GRU & 0.34 & 0.53 | 0.60 & 24 \\
GRU-large & 7.38 & 1.35 | 1.57 & 122 \\
\midrule
VQVAE-student & 0.34 & 1.35 | 1.76 & 24 \\
GRU-student & 0.34 & 0.56 | 0.59 & 22  \\
\bottomrule
\end{tabular}
\end{table}

\subsubsection{Completion}
\label{subsubsec:completion}
The main evaluation focuses on the signal completion, \ie, its reconstruction. We first introduce some details before presenting the results.

During testing, we use constraints on the samples similar to the data augmentation seen during training to prevent, to some extent, the performance degradation caused by a distribution shift. The main difference between both settings is the necessary insertion of a clear switch between masking and de-masking. Such \textit{edge} can be either \textit{rising} (0 to 1 switch) or \textit{falling} (1 to 0 switch). They are, indeed, necessary to precisely define the beginning of the update regime sequence (duration of $\approx1\,\text{s}$) and thus to compute the AUJ and PJ.

Moreover, we compute the metrics on small predicted sequences to reduce the number of comparisons with noisy ground-truth (detected by a threshold on inter-frame jerk).

Our completion evaluations are separated into two broad categories. First, (as usually done in the literature) we keep all hyperparameters constant to compare the variants against the baselines of both input configurations. Second, we observe the impact of modifying the mask hyperparameters, and more specifically the models' robustness to the maximal number of masked joints.

\para{Comprehensive metrics evaluation}
We exhaustively detail the quantitative results related to the main evaluation (\ie, constant hyper-parameters and metrics reported as $\text{mean}\pm \text{std}$) between configurations and baselines, which are presented in Tables~\ref{tab:fra_met} and \ref{tab:seq_met}.

First, the \emph{REACT} variants consistently outperform the evaluated
baselines on pose-level reconstruction metrics, with the largest
improvements observed on rotational errors. On update-regime metrics,
\texttt{RPM} remains particularly competitive and achieves the best
score on several criteria, while the strongest \emph{REACT} variants
remain comparatively close. These results highlight the same
reconstruction--stability trade-off discussed in the main paper.

Second, fusion ordering affects performance, although no ordering
dominates every criterion. Within the two-S4 family, the
Ref-last configuration performs best on most rotational and
update-regime metrics, whereas the Data-last configuration achieves
the lowest positional error. For the partitioned-S4 variants, we do
not observe a clear advantage of either initial spectral assignment,
suggesting that this initialization bias may be attenuated during
training. The consistently better results of the 2S4 variants further
suggest that partitioning a single S4 layer may constrain the available
modelling capacity.

Third, distillation substantially improves the students' rotational
accuracy while preserving their low latency and memory footprint.
However, these gains do not extend to positional accuracy or
update-regime stability, which remain below those of the structured
teacher.
This suggests that some benefits of the structured teacher may be
difficult to reproduce through output-level distillation alone.
As such, increasingly constrained architectures could be explored and compared, to provide a more comprehensive analysis of the benefits and limitations, and reach a better trade-off between latency and performance. 

Fourth, for GRU baselines, access to the raw-observation history
generally improves frame-based reconstruction metrics but degrades
update-regime stability. This suggests that recent observations provide
useful reconstruction cues, while unreliable observation history can
also destabilize regime transitions in architectures without explicit
reliability handling. The effect of removing the input projection
further indicates that the way both buffers are combined is itself an
important design choice.

Fifth, completion without the observation buffer remains competitive.
Although one-buffer models generally lose some reconstruction accuracy,
they can produce more stable updates, possibly because they are less
exposed to corrupted observation history. Choosing between both
formulations may therefore depend on the expected reliability of the
input signals and on the targeted reconstruction--stability trade-off.

\begin{figure}[p]
    \centering
    \includegraphics[width=0.85\linewidth]{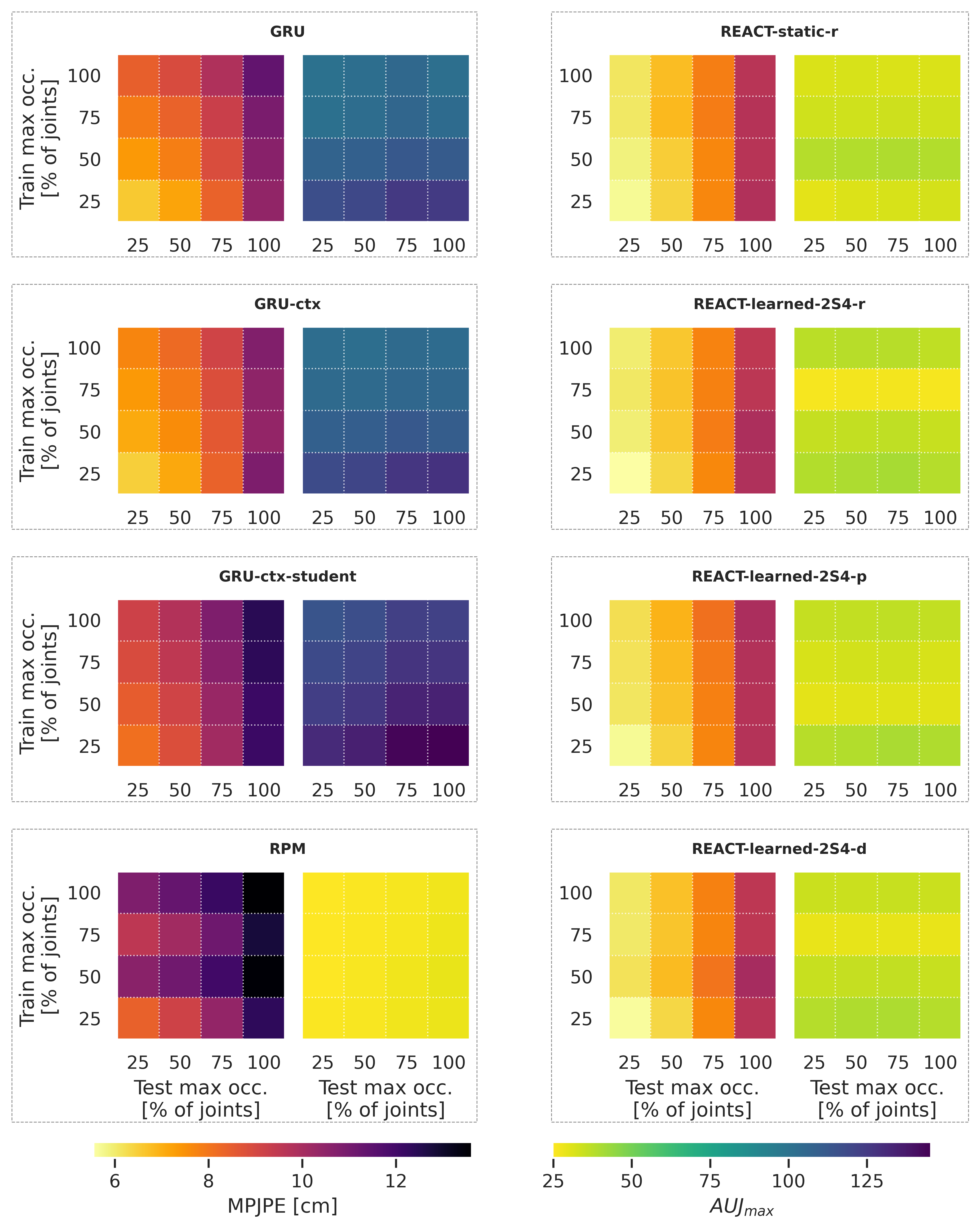}
    \caption{Robustness to masking severity. Rows correspond to $M_{\max}^{\text{train}} \in \{25,50,75,100\}\%$, and columns to $M_{\max}^{\text{test}} \in \{25,50,75,100\}\%$. Rows and columns respectively indicate the maximum masking ratio during training and testing. In both cases, $M_{\min}=0\%$. Each cell reports the mean $MPJPE$ (left) and $AUJ_{\max}=\max(AUJ_{T\to S}, AUJ_{S\to T})$ (right), where lower is better for both metrics (see colorbars). A larger $M_{\max}$ allows bigger joint masks, up to fully autoregressive segments.}
    \label{fig:maskrobustness}
\end{figure}

\para{Mask severity robustness}
As previously introduced, this specific evaluation quantifies the robustness to the mismatch between maximum masking range during training and testing. For each architecture, we train four checkpoints with $M_{\max}^{\text{train}} \in \{25,50,75,100\}\%$ while keeping $M_{\min}=0\%$ and evaluate them under four regimes $M_{\max}^{\text{test}} \in \{25,50,75,100\}\%$. This yields 16 values per model and for each metric. We further reduce the number of objective criteria to 2 (for an interpretable comparison): the MPJPE, and $AUJ_{\max}=\max(AUJ_{S\to T}, AUJ_{T\to S})$ (both are metrics where lower is better). Although incomplete, these metrics provide an overview of the evolution of both the pose reconstruction and the update regime dynamics.
Note that during testing, we remove the joint mask from some subsequences to compute the update regime metrics. The results of our evaluation are detailed in Figure~\ref{fig:maskrobustness}.

As hinted in the previous evaluation, GRU-based architectures provide relevant positional error at the cost of a jerky motion during transitions. Furthermore, their performance on positional metrics consistently degrades with an increase in mask severity. The additional buffer between \texttt{GRU} and \texttt{GRU-ctx} may improve the robustness to the mask severity, although it remains limited. The similar patterns exhibited by \texttt{GRU-ctx-student}, trained by distillation, may suggest that the masking range impact depends at least partly on the underlying architecture.

\texttt{RPM} prioritizes motion smoothness and is consistent on the related metric with a virtually constant $AUJ_{max}$ no matter the mask severity. However, its positional error suffers from possible under-correction and seem to increase when trained with more occluded signals.

Finally, our variants provide both a good trade-off between motion smoothness and low positional error, as well as a reduced vulnerability to the training mask severity. The similar sensitivity patterns across \emph{REACT} variants suggest that their shared system structure may contribute more strongly than the
specific fusion ordering or other minor modifications to this behaviour.

\begin{landscape}

\begin{table}[p]
\centering
\small
\setlength{\tabcolsep}{5pt}
\renewcommand{\arraystretch}{1.2}
\caption{Pose-level metrics}
\label{tab:fra_met}

\begin{tabular}{lcccccc}
\toprule
Model & \makecell[c]{MPJPE \\ \text{[cm]}} ($\downarrow$) & \makecell[c]{P-MPJPE \\ \text{[cm]}} ($\downarrow$) & \makecell[c]{MPJLAE \\ \text{[deg]}} ($\downarrow$) & \makecell[c]{MPJAE \\ \text{[deg]}} ($\downarrow$) & \makecell[c]{MPJPVE \\ \text{[cm/s]}} ($\downarrow$) & \makecell[c]{MPJAVE \\ \text{[deg/s]}} ($\downarrow$) \\
\midrule
\midrule

RPM  & $11.12 \pm 6.89$ & $8.71 \pm 4.05$ & $24.18 \pm 4.67$ & $30.81 \pm 18.57$ & $48.46 \pm 64.06$ & $73.12 \pm 163.20$ \\
\midrule
GRU-ctx & $8.83 \pm 6.39$ & $6.60 \pm 4.14$ & $22.69 \pm 5.52$ & $27.04 \pm 15.91$ & $52.88 \pm 55.49$ & $82.41 \pm 144.95$ \\

GRU-ctx-large & $11.04 \pm 6.73$ & $8.35 \pm 3.95$ & $20.62 \pm 5.05$ & $28.29 \pm 19.73$ & $56.40 \pm 64.55$ & $82.43 \pm 205.53$ \\

GRU-ctx-large-noproj & $16.11 \pm 8.81$ & $11.72 \pm 4.58$ & $77.96 \pm 15.41$ & $65.45 \pm 17.43$ & $50.01 \pm 45.98$ & $75.64 \pm 93.31$ \\
\midrule

GRU-ctx-student & $10.60 \pm 7.01$ & $7.97 \pm 4.12$ & $14.41 \pm 5.61$ & $24.79 \pm 20.42$ & $58.08 \pm 67.59$ & $85.83 \pm 195.37$ \\

VQVAE-ctx-student & $10.89 \pm 7.36$ & $8.47 \pm 4.58$ & $13.91 \pm 6.08$ & $23.94 \pm 21.29$ & $63.01 \pm 78.62$ & $96.40 \pm 223.45$ \\

\midrule
REACT-static-lfp\_hfd-r      & $7.85 \pm 6.57$  & $5.93 \pm 4.28$  & $\msecond{12.32 \pm 6.93}$ & $20.29 \pm 17.02$ & $44.72 \pm 46.91$ & $63.93 \pm 81.17$ \\
REACT-static-lfp\_hfd-p       & $7.89 \pm 6.22$  & $6.05 \pm 4.18$  & $14.12 \pm 6.28$ & $20.40 \pm 15.08$ & $43.57 \pm 46.94$ & $\msecond{62.13 \pm 92.82}$ \\
REACT-static-lfp\_hfd-d       & $7.93 \pm 6.33$  & $6.11 \pm 4.20$  & $13.24 \pm 6.23$ & $20.41 \pm 16.00$ & $43.44 \pm 45.73$ & $62.97 \pm 86.96$ \\

\midrule
REACT-learned-lfp\_hfd-r     & $7.88 \pm 6.50$  & $5.99 \pm 4.25$  & $13.06 \pm 6.38$ & $20.32 \pm 16.51$ & $45.94 \pm 49.49$ & $64.92 \pm 102.24$ \\
REACT-learned-lfp\_hfd-p      & $8.08 \pm 6.28$  & $6.21 \pm 4.20$  & $14.29 \pm 6.49$ & $20.69 \pm 15.09$ & $45.09 \pm 47.99$ & $65.07 \pm 96.04$ \\
REACT-learned-lfp\_hfd-d      & $7.92 \pm 6.45$  & $6.08 \pm 4.27$  & $13.76 \pm 6.54$ & $20.23 \pm 16.40$ & $45.20 \pm 49.05$ & $65.26 \pm 107.02$ \\

\midrule
REACT-static-lfd\_hfp-r      & $\msecond{7.71 \pm 6.45}$ & $\mbest{5.82 \pm 4.29}$ & $12.65 \pm 6.54$ & $\msecond{20.02 \pm 16.71}$ & $45.39 \pm 47.91$ & $65.93 \pm 98.62$\\
REACT-static-lfd\_hfp-p    & $7.96 \pm 6.38$ & $6.14 \pm 4.29$ & $14.62 \pm 6.47$ & $21.13 \pm 14.97$ & $45.82 \pm 49.20$ & $67.07 \pm 102.71$ \\
REACT-static-lfd\_hfp-d    & $8.00 \pm 6.39$ & $6.18 \pm 4.26$ & $15.12 \pm 7.66$ & $21.52 \pm 15.96$ & $44.77 \pm 45.41$ & $67.71 \pm 87.70$ \\

\midrule
REACT-learned-lfd\_hfp-r     & $7.90 \pm 6.60$ & $6.06 \pm 4.27$ & $12.93 \pm 6.67$ & $20.23 \pm 16.89$ & $48.01 \pm 60.52$ & $67.25 \pm 121.30$ \\
REACT-learned-lfd\_hfp-p     & $8.15 \pm 6.37$ & $6.30 \pm 4.31$ & $14.91 \pm 6.51$ & $21.56 \pm 15.02$ & $44.77 \pm 47.46$ & $64.11 \pm 86.83$  \\
REACT-learned-lfd\_hfp-d      & $8.09 \pm 6.28$ & $6.34 \pm 4.26$ & $14.30 \pm 6.29$ & $21.09 \pm 15.08$ & $44.39 \pm 46.80$ & $64.18 \pm 94.70$ \\

\midrule
REACT-learned-2S4-r& $7.75 \pm 6.29$ & $\msecond{5.87 \pm 4.20}$ & $\mbest{12.21 \pm 6.56}$ & $\mbest{19.62 \pm 15.41}$ & $\mbest{42.35 \pm 43.97}$ & $\mbest{60.21 \pm 76.35}$ \\
REACT-learned-2S4-p & $7.92 \pm 6.50$ & $5.96 \pm 4.22$ & $14.85 \pm 6.70$ & $20.63 \pm 17.11$ & $44.43 \pm 48.46$ & $65.66 \pm 100.84$ \\
REACT-learned-2S4-d & $\mbest{7.70 \pm 6.36}$ & $5.88 \pm 4.17$ & $13.84 \pm 6.65$ & $20.10 \pm 15.47$ & $\msecond{42.83 \pm 43.78}$ & $62.66 \pm 74.68$ \\
\midrule
\midrule
GRU    & $9.26 \pm 6.45$  & $6.86 \pm 4.05$  & $23.68 \pm 5.11$ & $27.80 \pm 16.59$ & $52.54 \pm 55.94$ & $80.25 \pm 144.00$ \\

GRU-large & $12.91 \pm 7.70$ & $9.52 \pm 4.16$ & $22.03 \pm 4.72$ & $31.13 \pm 20.10$ & $54.07 \pm 62.64$ & $78.64 \pm 198.79$ \\

\midrule

GRU-student & $10.71 \pm 6.97$ & $8.07 \pm 4.10$ & $14.42 \pm 5.41$ & $24.81 \pm 20.34$ & $57.62 \pm 67.62$ & $85.09 \pm 193.00$ \\

VQVAE-student & $10.75 \pm 7.24$ & $8.32 \pm 4.57$ & $14.15 \pm 6.01$ & $24.06 \pm 21.24$ & $64.43 \pm 79.06$ & $97.41 \pm 223.76$ \\
\bottomrule
\end{tabular}

\end{table}

\end{landscape}


\begin{landscape}

\begin{table}[p]
\centering
\small
\setlength{\tabcolsep}{5pt}
\renewcommand{\arraystretch}{1.2}

\caption{Sequence- and event-based metrics}
\label{tab:seq_met}

\begin{tabular}{lcccccc}
\toprule
Model & PJ$_{T\to S}$ ($\downarrow$) & PJ$_{S\to T}$ ($\downarrow$) &
AUJ$_{T\to S}$ ($\downarrow$) & AUJ$_{S\to T}$ ($\downarrow$)  & TPSE & RCS \\
\midrule
\midrule

RPM  & $\msecond{1.47 \pm 5.55}$ & $\mbest{1.80 \pm 7.63}$ & $\mbest{25.36 \pm 13.07}$ & $\mbest{26.40 \pm 22.30}$ & $0.40 \pm 0.08$ & $1128.99 \pm 285.91$\\

\midrule

GRU-ctx & $5.17 \pm 3.60$ & $8.65 \pm 6.25$ & $105.27 \pm 47.32$ & $105.25 \pm 46.11$ & $0.60 \pm 0.12$ & $1137.50 \pm 287.55$ \\

GRU-ctx-large & $6.41 \pm 3.93$ & $10.02 \pm 6.81$ & $134.44 \pm 70.58$ & $135.31 \pm 63.87$ & $0.59 \pm 0.11$ & $1132.83 \pm 288.30$ \\

GRU-ctx-large-noproj & $3.15 \pm 2.17$ & $4.50 \pm 2.80$ & $43.99 \pm 25.41$ & $47.72 \pm 21.12$ & $0.48 \pm 0.12$ & $1141.30 \pm 284.59$\\
\midrule

GRU-ctx-student & $5.62 \pm 4.73$ & $11.59 \pm 7.80$ & $104.68 \pm 48.22$ & $127.07 \pm 53.78$ & $0.59 \pm 0.12$ & $1124.68 \pm 290.83$ \\

VQVAE-ctx-student & $6.64 \pm 5.29$ & $14.17 \pm 9.64$ & $128.40 \pm 57.70$ & $148.76 \pm 63.57$ & \msecond{$0.63 \pm 0.11$} & $1114.12 \pm 300.69$ \\

\midrule
REACT-static-lfp\_hfd-r      & $2.38 \pm 2.95$ & $3.32 \pm 4.22$ & $33.80 \pm 12.96$ & $33.79 \pm 14.23$ & $0.49 \pm 0.09$ & $1136.50 \pm 284.67$\\
REACT-static-lfp\_hfd-p      & $2.01 \pm 3.73$ & $2.92 \pm 5.15$ & $28.82 \pm 12.95$ & $30.59 \pm 22.61$ & $0.48 \pm 0.10$ & $1134.93 \pm 287.78$\\
REACT-static-lfp\_hfd-d      & $2.05 \pm 2.37$ & $3.05 \pm 3.73$ & $29.50 \pm 10.20$ & $31.32 \pm 13.17$ & $0.48 \pm 0.09$ & $1133.27 \pm 287.62$\\

\midrule
REACT-learned-lfp\_hfd-r      & $2.03 \pm 3.24$ & $2.68 \pm 4.75$ & $28.38 \pm 10.43$ & $30.18 \pm 17.22$ & $0.49 \pm 0.09$ & $1134.94 \pm 286.72$\\
REACT-learned-lfp\_hfd-p      & $2.14 \pm 2.94$ & $3.87 \pm 4.97$ & $31.78 \pm 30.71$ & $33.51 \pm 18.78$ & $0.48 \pm 0.09$ & $\msecond{1139.36 \pm 284.16}$\\
REACT-learned-lfp\_hfd-d      & $2.30 \pm 2.77$ & $3.67 \pm 4.48$ & $33.46 \pm 10.85$ & $34.21 \pm 13.33$ & $0.49 \pm 0.10$ & $1136.04 \pm 283.83$\\

\midrule
REACT-static-lfd\_hfp-r      & $2.47 \pm 2.38$ & $3.35 \pm 4.07$ & $34.70 \pm 12.65$ & $34.71 \pm 14.26$ & $0.49 \pm 0.09$ & $1136.25 \pm 282.86$ \\
REACT-static-lfd\_hfp-p    & $2.65 \pm 3.97$ & $3.93 \pm 5.09$ & $35.47 \pm 22.25$ & $35.80 \pm 19.58$ & $0.49 \pm 0.09$ & $1135.47 \pm 287.92$ \\
REACT-static-lfd\_hfp-d    & $2.12 \pm 2.75$ & $2.90 \pm 3.38$ & $30.43 \pm 11.41$ & $31.82 \pm 11.99$ & $0.49 \pm 0.09$ & $1136.93 \pm 286.39$ \\

\midrule
REACT-learned-lfd\_hfp-r     & $2.47 \pm 5.20$ & $3.27 \pm 6.78$ & $38.75 \pm 187.62$ & $39.72 \pm 216.88$ & $0.51 \pm 0.10$ & $\mbest{1143.14 \pm 282.43}$ \\
REACT-learned-lfd\_hfp-p     & $1.99 \pm 1.75$ & $3.44 \pm 4.29$ & $30.67 \pm 8.08$ & $31.77 \pm 12.56$ & $0.48 \pm 0.09$ & $1133.69 \pm 283.79$ \\
REACT-learned-lfd\_hfp-d      & $2.13 \pm 2.30$ & $3.25 \pm 4.28$ & $31.68 \pm 9.01$ & $32.59 \pm 13.68$ & $0.48 \pm 0.09$ & $1134.15 \pm 286.08$ \\

\midrule
REACT-learned-2S4-r& $\mbest{1.41 \pm 1.94}$ & $\msecond{2.02 \pm 3.42}$ & $\msecond{25.57 \pm 5.69}$ & $\msecond{26.75 \pm 10.30}$ & $0.47 \pm 0.09$ & $1134.79 \pm 283.86$\\
REACT-learned-2S4-p & $2.08 \pm 2.84$ & $3.80 \pm 5.03$ & $29.94 \pm 9.97$ & $33.29 \pm 18.70$ & $0.48 \pm 0.09$ & $1134.90 \pm 286.26$\\
REACT-learned-2S4-d & $1.68 \pm 1.50$ & $2.71 \pm 2.67$ & $27.40 \pm 6.68$ & $29.36 \pm 9.07$ & $0.48 \pm 0.09$ & $1135.07 \pm 286.32$\\

\midrule
\midrule

GRU            & $4.87 \pm 3.43$ & $8.90 \pm 6.78$ & $98.98 \pm 43.57$ & $105.38 \pm 45.13$ & $0.59 \pm 0.12$ & $1138.25 \pm 288.10$\\

GRU-large & $4.80 \pm 4.19$ & $8.64 \pm 6.48$ & $78.46 \pm 45.81$ & $92.57 \pm 46.63$ & $0.56 \pm 0.12$ & $1126.42 \pm 286.82$ \\
\midrule

GRU-student & $5.58 \pm 4.64$ & $11.60 \pm 7.75$ & $104.03 \pm 47.40$ & $126.89 \pm 53.54$ & $0.59 \pm 0.12$ & $1124.47 \pm 288.47$ \\

VQVAE-student & $6.84 \pm 5.31$ & $14.07 \pm 9.51$ & $133.09 \pm 59.83$ & $152.04 \pm 64.68$ & $\mbest{0.63 \pm 0.11}$ & $1118.94 \pm 299.94$ \\

\bottomrule
\end{tabular}

\end{table}

\end{landscape}

\section{REACT internal analysis}
\label{subsec:supp_internal}
\subsection{Objective}
In this section, we focus on a deeper analysis of the internal behaviour of our models (\texttt{REACT-learned-2S4-r}, \texttt{REACT-learned-2S4-p} and \texttt{REACT-learned-2S4-d}) with a particular focus on the fusion module, depending on the per-joint and per-frame ground-truth mask (``Observable'' = ``Obs'' or ``Occluded'' = ``Occ''). 
First, we represent the conditional distributions of several fusion signals, such as the relative weights of the branches in the final mix (such that $\sum_{i \in \{Phys, Data, Ref\}} w_i = 1$) and how they change with the mask and the variant (\texttt{-r}, \texttt{-d} and \texttt{-p}).
Then, we focus on a single variant (\texttt{REACT-learned-2S4-r}) to a) estimate the performance of occlusion classification and b) observe the effect of the defiance on the update amplitude bottleneck.

\subsection{Protocol and criteria}
For each joint~$j$, let $T_j^s$ denote the set of frames belonging to
scenario $s\in\{\mathrm{Occ},\mathrm{Obs}\}$. We define the
candidate-wise winner indicator as
\begin{equation}
    I_{i,j,t}
    =
    \mathds{1}\!\left[
        i =
        \operatorname*{arg\,max}_{k\in
        \{\mathrm{Phys},\mathrm{Data},\mathrm{Ref}\}}
        w_{k,j,t}
    \right],
\end{equation}
where $w_{k,j,t}$ is the effective fusion weight assigned to candidate
$k$. The conditional winner rate is then:
\begin{equation}
    \mathrm{WR}_{i,j}^{s}
    =
    \frac{1}{|T_j^s|}
    \sum_{t\in T_j^s} I_{i,j,t}.
\end{equation}
It is the fraction of frames where a candidate has the largest weight. Although useful to have a broad idea of the weighting mix, the winner rate ignores near-ties and thus discards interesting nuances in the distribution. Thus, we add the per-frame normalized entropy of the weights distribution:
\begin{equation}
    H_{j,t}
    =
    -\frac{1}{\log 3}
    \sum_{i\in\{\mathrm{Phys},\mathrm{Data},\mathrm{Ref}\}}
    w_{i,j,t}\log w_{i,j,t},
\end{equation} which leads to the conditional entropy:
\begin{equation}
    H_j^s
    =
    \frac{1}{|T_j^s|}
    \sum_{t\in T_j^s} H_{j,t}.
\end{equation}
We further leverage the Balanced Accuracy, a score to estimate the performance of a binary classifier under possibly unbalanced classes, defined as: $\text{Balanced Accuracy} = \frac{TPR + TNR}{2}$, with $TPR=\frac{TP}{TP+FN}$, $TNR=\frac{TN}{TN+FP}$. Finally, we display the per-joint local angular error (\ie, angular error between ground-truth and prediction as used in the $\text{MPJLAE}$ but without aggregating over joints).

\subsection{Results}
Here, we present several evaluations of our models on the test set. We articulate the presentation around the candidate weighting, an estimation of the internal classification performance, and a typical effect of our defiance mechanism on the output amplitude.

\para{Candidate weighting}
The branch-wise conditional winner rate is presented in Table~\ref{tab:wr}. First, we can observe that a candidate has a higher winner rate when it is the last interpolated one than when it is not. E.g., Ref has a winner rate of $0.60$ in Ref-last variant but only $0.37$ and $0.05$ in Data-last and Phys-last. Such an observation is consistent with the implicit bias of the hierarchical interpolation. Second, the Ref contribution systematically decreases between Obs and Occ scenarios, while it increases for Data. Overall, Phys is also more solicited in Occ, except in \texttt{-p} where its winner rate slightly decreases while staying dominant. This is consistent with the intended prior: under occlusion, direct observations become unreliable and the fusion
therefore shifts weight away from Ref toward predictive
candidates. Third, Phys seems to be preferred by the system in most cases (in $2/3$ cases under Obs and $3/3$ under Occ), which is consistent with its intended
role as the reactive short-horizon candidate. Fourth, the standard deviation is often large (\ie, $1/4$ to $1/3$), which indicates a large per-joint diversity in the results.
In Table~\ref{tab:entropy_agg}, we show an entropy averaged over joints to have an idea of the fusion decisiveness in the various scenarios. Broadly, the entropy is lower under occlusion and thus the fusion tends to be more decisive, which might be linked to the progressive disappearance of Ref as hinted in Table~\ref{tab:wr}. The \texttt{-p} variant also exhibits lower entropy, coinciding with
the substantially larger winner rate of \emph{Phys}.

\begin{figure}[t]
    \centering
    \includegraphics[width=1\linewidth]{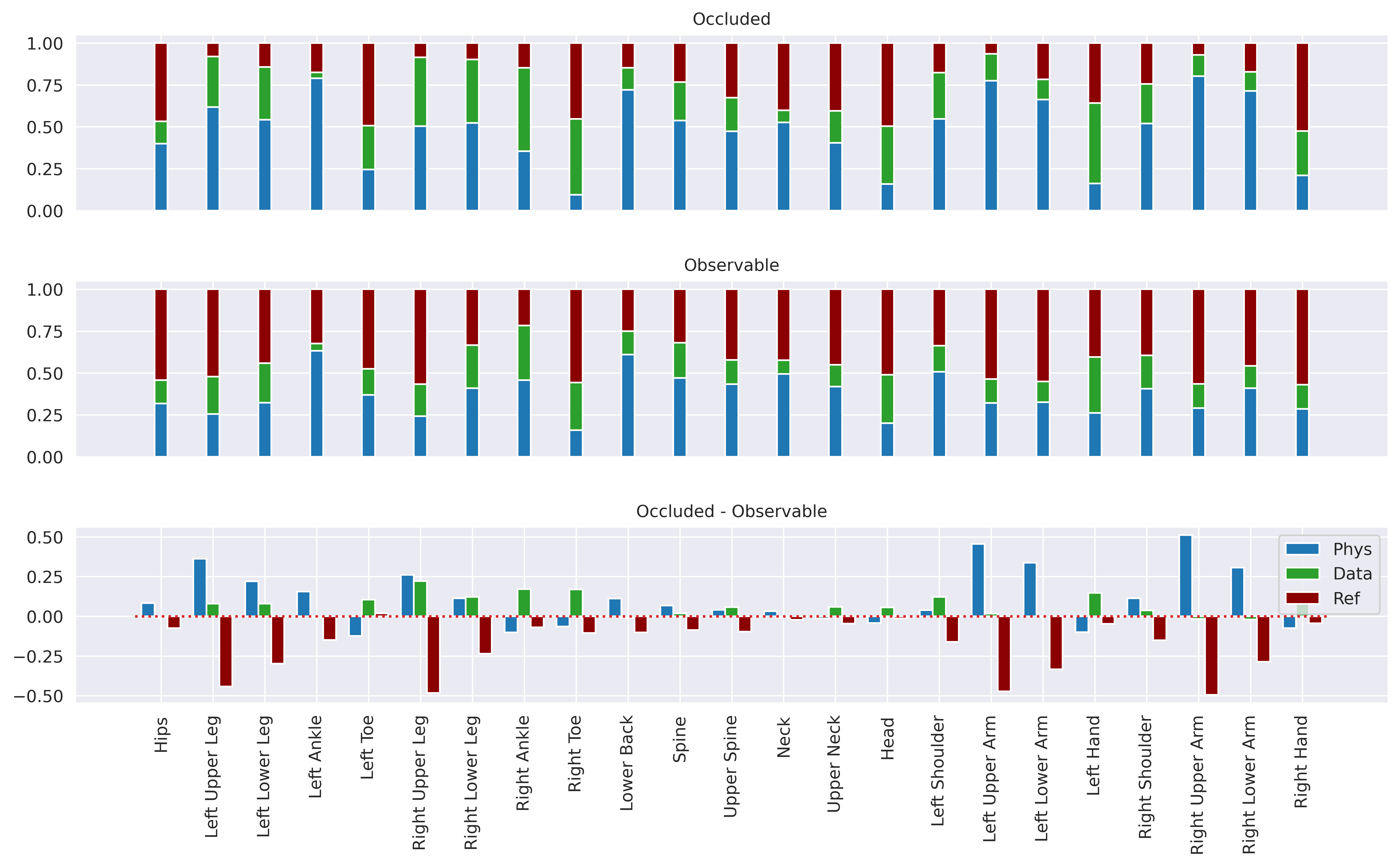}
    \caption{Per-joint average fusion weights for the Ref-last (\mbox{\texttt{-r}}) variant under ``Occluded'' (top) and ``Observable'' (middle) conditions. The scenario difference is presented in ``Occluded--Observable'' (bottom), where positive values indicate an increased contribution under occlusion.}
    \label{fig:mix_occobs}
\end{figure}

We analyse in finer detail the spatial distribution of the weights in Figure~\ref{fig:mix_occobs}. Here, we present the \texttt{-r} variant but the distributions were similar for \texttt{-d} and \texttt{-p}. In the ``Observable'' scenario we have a majority of Ref, which is partly caused by the branch advantageous weighting in this variant but also by its performance (\ie, a skip connection). However, in ``Occluded'' Ref is noticeably less used, which is apparent in its negative values in the ``Occluded - Observable'' plot, while Phys gains more traction. Although this information is already hinted by the candidates' winner rate, we observe here a disparity of the variations across joints. For instance, axial joints (\eg, hips, spine) have a comparatively lower change in the average weights mixing between both scenarios. A choice between reactive and conservative choices for these joints might be more difficult, since such joints are structural for the overall motion. A possible explanation is that the kinematic hierarchy leads the models to encounter diverging objectives during training, \ie, between the need for stability of such important joints (including under occlusion) and an explosion of the error if they are over-smoothed. Another possible cause is a less accurate classification of the occlusion scenario for these joints, which mechanically leads to an absence of behavioural shift. The binary classification evaluation is addressed later. In contrast, some limb joints such as the shoulders and legs have a clearly distinct weight distribution between both cases, with a particular preference for the Phys branch. Finally, the end-effectors (feet/hands) only give more weight to the Data branch under occlusion, and not under observation. This last result might suggest a) an advantage of a long-horizon prediction context or b) a potentially over-conservative strategy of Phys for such dynamic joints. %
As hinted previously, beyond structural difficulties a possible explanation for the differences in weight distribution might also be an unequal performance of the occlusion classification.

\begin{table}[t]
\caption{Aggregated ($mean \pm std$ across joints) winner rate by branch, last interpolated source, and occlusion scenario (Obs = Observable, Occ = Occluded).}
\label{tab:wr}
\centering
\small
\setlength{\tabcolsep}{4pt}
\begin{tabular}{llccc}
\toprule
Candidate & Scenario & Ref-last (-r) & Data-last (-d) & Phys-last (-p) \\
\midrule
\multirow{2}{*}{Ref}
 & Obs & $0.60 \pm 0.35$ & $0.37 \pm 0.35$ & $0.05 \pm 0.12$ \\
 & Occ & $0.23 \pm 0.35$ & $0.10 \pm 0.19$ & $0.02 \pm 0.08$ \\
\midrule
\multirow{2}{*}{Data}
 & Obs & $0.04 \pm 0.07$ & $0.10 \pm 0.19$ & $0.06 \pm 0.14$ \\
 & Occ & $0.14 \pm 0.27$ & $0.20 \pm 0.27$ & $0.12 \pm 0.27$ \\
\midrule
\multirow{2}{*}{Phys}
 & Obs & $0.36 \pm 0.35$ & $0.53 \pm 0.35$ & $0.89 \pm 0.17$ \\
 & Occ & $0.63 \pm 0.41$ & $0.71 \pm 0.31$ & $0.85 \pm 0.27$ \\
\bottomrule
\end{tabular}
\end{table}

\begin{table}[ht!]
\caption{Aggregated and normalized ($mean \pm std$ across joints, values in $[0,1]$) fusion weight entropy under observable and occluded conditions. Lower values indicate a more decisive fusion. Maximal value (1) corresponds to the three branches sharing the same value of $1/3$.}
\label{tab:entropy_agg}
\centering
\small
\setlength{\tabcolsep}{8pt}
\begin{tabular}{lccc}
\toprule
Scenario & Ref-last (\texttt{-r}) & Data-last (\texttt{-d}) & Phys-last (\texttt{-p}) \\
\midrule
Obs & $0.87 \pm 0.06$ & $0.88 \pm 0.05$ & $0.78 \pm 0.10$ \\
Occ & $0.80 \pm 0.12$ & $0.81 \pm 0.09$ & $0.65 \pm 0.16$ \\
\bottomrule
\end{tabular}

\end{table}

\begin{figure}[p]
    \centering
    \includegraphics[width=0.8\linewidth]{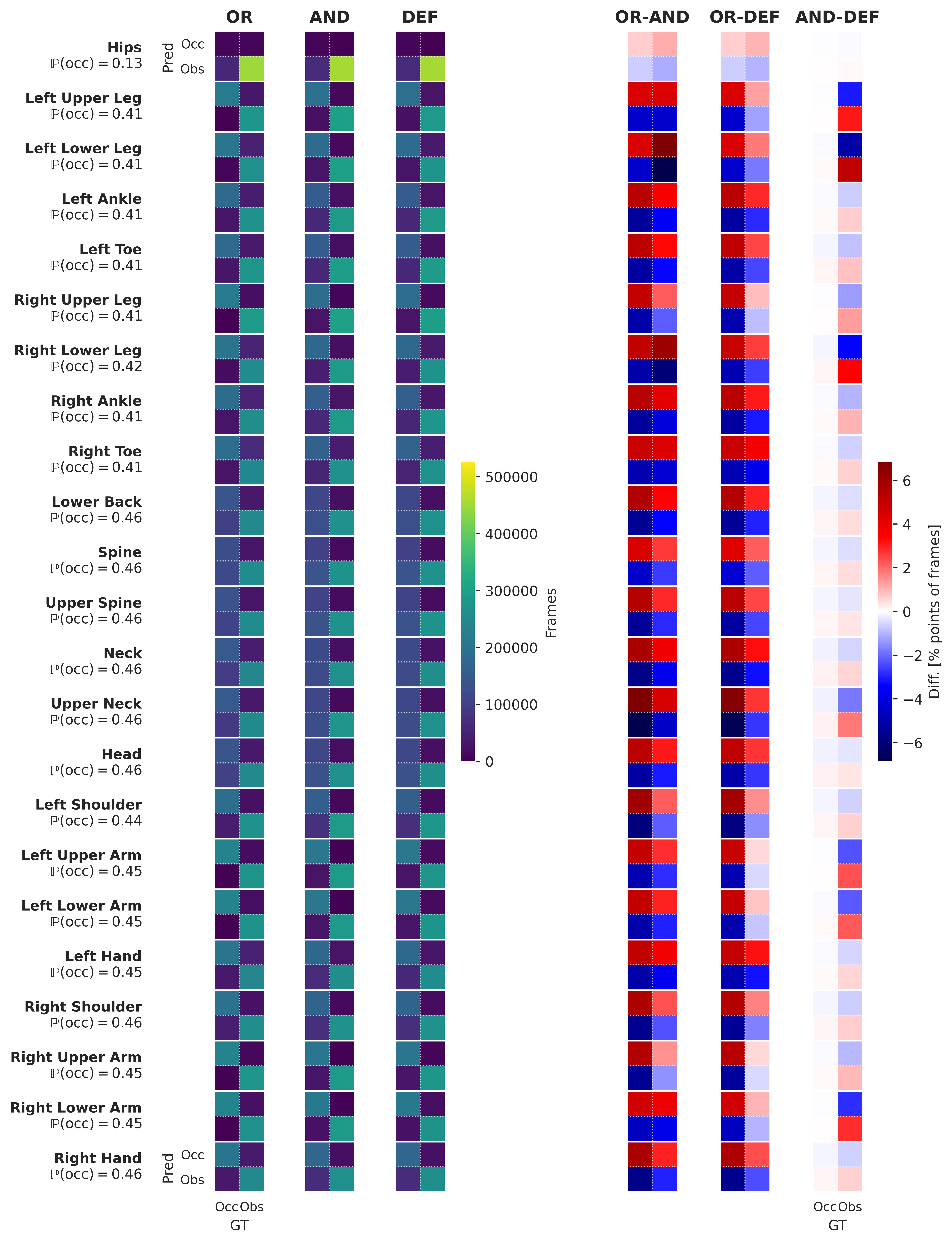}
    \caption{Per-joint occlusion classification confusion matrices. In each $2\times2$ matrix, cells correspond to (top-left) TP $(\text{Pred}=\text{Occ},\,\text{GT}=\text{Occ})$, (top-right) FP $(\text{Pred}=\text{Occ},\,\text{GT}=\text{Obs})$, (bottom-left) FN $(\text{Pred}=\text{Obs},\,\text{GT}=\text{Occ})$, and (bottom-right) TN $(\text{Pred}=\text{Obs},\,\text{GT}=\text{Obs})$. On the left part of the figure, the cells are coloured \wrt the raw frame counts (\ie, also reflects bias towards predominant Obs class). On its right part, we show the pairwise difference between decision rules (OR--AND, OR--DEF, AND--DEF) as percentage points of frames (normalized by total frames). A red case means the left operand has more frames in this class than the right operand (increases in TP/TN is better, while increases in FP/FN is worse).}
    \label{fig:confmat}
\end{figure}

\para{Reliability estimation}
As a reminder, the model's internal classification and priors are mainly articulated around two signals. There is first the instantaneous and frame-by-frame prediction of the network, which is by design reactive but also unstable (prone to jump between states). Second, we further rely on the defiance buffer, which acts as both a hysteresis and a history. Of course, both signals are not a binary switch but continuous values in $[0,1]$. This allows us to treat the instantaneous signal as a probability of occlusion (more robust and informative) and to use it to increment/decrement the defiance buffer asymmetrically (\eg, slower increase than decrease to reduce FP) and conditionally. This non-linear integration of the class estimation signals to our model might lead to minor differences \wrt a binary classification. Moreover, final defiance signals are processed by MLPs to modulate the effect of the class estimation and model uncertainty for each joint.
However, first we propose to estimate the performance of our classification. Because the continuous reliability signals are subsequently transformed by the fusion module, no single binary rule exactly reproduces their effective use. We therefore evaluate three simple proxy decision rules obtained by thresholding the instantaneous classifier and normalized defiance state at $0.5$. We have two natural choices to fuse these boolean values, either using ``OR'' or ``AND''. I.e., let $\hat{p}_{tj},\hat{d}_{tj}\in [0,1]$ be respectively the instantaneous frame class probability and the normalized defiance buffer value (with $t$ the frame and $j$ the joint considered). Then $\hat{\text{occ}}^{\text{AND}}_{tj} = \mathds{1}[\hat{p}_{tj}>0.5 \wedge \hat{d}_{tj}>0.5]$ and $\hat{\text{occ}}^{\text{OR}}_{tj} = \mathds{1}[\hat{p}_{tj}>0.5 \vee \hat{d}_{tj}>0.5]$. To observe the evolution of the defiance buffer (which is directly piloted by the frame classifier), we also pose $\hat{\text{occ}}^{\text{DEF}}_{tj} = \mathds{1}[\hat{d}_{tj}>0.5]$. These three decision rules are respectively denoted ``AND'', ``OR'' and ``DEF''. Altogether, these rules provide three interpretable binary views of the continuous reliability signals used by the model. We present the results as per-joint occlusion confusion matrices and their differences in Figure~\ref{fig:confmat}. To consider the class imbalance, we further display the probability of occlusion $\mathbb{P}(\text{occ})$. The disparity between occlusion probability across joints is mostly driven by the per-body-parts masking. Indeed, in this scenario we remove ``Hips'' from the set of joints that can be masked independently (although it can still happen during body-part-level masking). Initially, the objective was to both reflect the necessary protection of a crucial joint in a local representation and ensure training stability. We kept the same distribution at test time for consistency. With the exception of the root, most joints have an almost balanced testing set ($>40\%$ of Occ). On the results themselves, we can first observe that despite the imbalance, hips exhibits a low TP rate and a substantial amount of FN. For most other joints, TP and TN appear to dominate FP and FN, which suggests that our classifiers perform better than chance.
However, the axial joints (spine, head, \etc) exhibit a noteworthy high false-negative rate (Occluded predicted as Observable). The pairwise difference between decision rules shows that TP and FP tend to increase in the ``OR'' case (which is consistent with giving more weight to the instantaneous and reactive frame classification). In contrast, ``AND'' and ``DEF'' are less prone to predict an Occ class. Their own comparison shows that ``DEF'' and ``AND'' are quite similar on TP/FN but ``AND'' is better in terms of TN/FP.

\begin{table}[t]
\caption{Per-joint balanced accuracy of three proxy decision rules
derived from the model's continuous reliability signals.}
\label{tab:classifiers}
\centering
\small
\setlength{\tabcolsep}{3pt}
\begin{tabular}{lccc}

\toprule
Joint & \makecell[c]{OR} & \makecell[c]{AND} & DEF \\
\midrule
Hips - $\mathbb{P}(\text{Occ})=0.13$& $0.57$ & $0.55$& $0.55$\\
Left Upper Leg - $\mathbb{P}(\text{Occ})=0.41$ & $0.94$ &$0.93$ &$0.90$ \\
Left Lower Leg - $\mathbb{P}(\text{Occ})=0.41$& $0.91$&$0.91$ &$0.87$ \\
Left Ankle - $\mathbb{P}(\text{Occ})=0.41$&$0.85$ &$0.82$ & $0.81$\\
Left Toe - $\mathbb{P}(\text{Occ})=0.41$& $0.86$& $0.83$& $0.82$ \\
Right Upper Leg - $\mathbb{P}(\text{Occ})=0.41$& $0.96$&$0.92$ &$0.91$ \\
Right Lower Leg - $\mathbb{P}(\text{Occ})=0.42$&$0.88$ &$0.87$ & $0.84$\\
Right Ankle - $\mathbb{P}(\text{Occ})=0.41$& $0.85$&$0.82$ & $0.81$\\
Right Toe - $\mathbb{P}(\text{Occ})=0.41$&$0.83$ &$0.81$ & $0.81$\\
Lower Back - $\mathbb{P}(\text{Occ})=0.46$&$0.73$ & $0.70$&$0.70$ \\
Spine - $\mathbb{P}(\text{Occ})=0.46$&$0.71$ & $0.69$& $0.68$\\
Upper Spine - $\mathbb{P}(\text{Occ})=0.46$&$0.73$ &$0.69$ &$0.69$ \\
Neck - $\mathbb{P}(\text{Occ})=0.46$&$0.73$ & $0.71$& $0.70$\\
Upper Neck - $\mathbb{P}(\text{Occ})=0.46$& $0.75$& $0.72$&$0.70$ \\
Head - $\mathbb{P}(\text{Occ})=0.46$&$0.72$ &$0.69$ & $0.69$\\
Left Shoulder - $\mathbb{P}(\text{Occ})=0.44$&$0.87$ &$0.82$ &$0.82$ \\
Left Upper Arm - $\mathbb{P}(\text{Occ})=0.45$&$0.97$ & $0.94$&$0.92$ \\
Left Lower Arm - $\mathbb{P}(\text{Occ})=0.45$& $0.96$& $0.93$& $0.91$\\
Left Hand - $\mathbb{P}(\text{Occ})=0.45$&$0.84$ &$0.82$ & $0.82$\\
Right Shoulder - $\mathbb{P}(\text{Occ})=0.46$&$0.86$ &$0.82$ & $0.82$\\
Right Upper Arm - $\mathbb{P}(\text{Occ})=0.45$&$0.97$ &$0.93$ & $0.92$\\
Right Lower Arm - $\mathbb{P}(\text{Occ})=0.45$&$0.96$ &$0.94$ & $0.91$\\
Right Hand - $\mathbb{P}(\text{Occ})=0.46$&$0.86$ &$0.82$ & $0.82$\\
\bottomrule
\end{tabular}
\end{table}

\begin{figure}[t]
    \centering
    \includegraphics[width=1\linewidth]{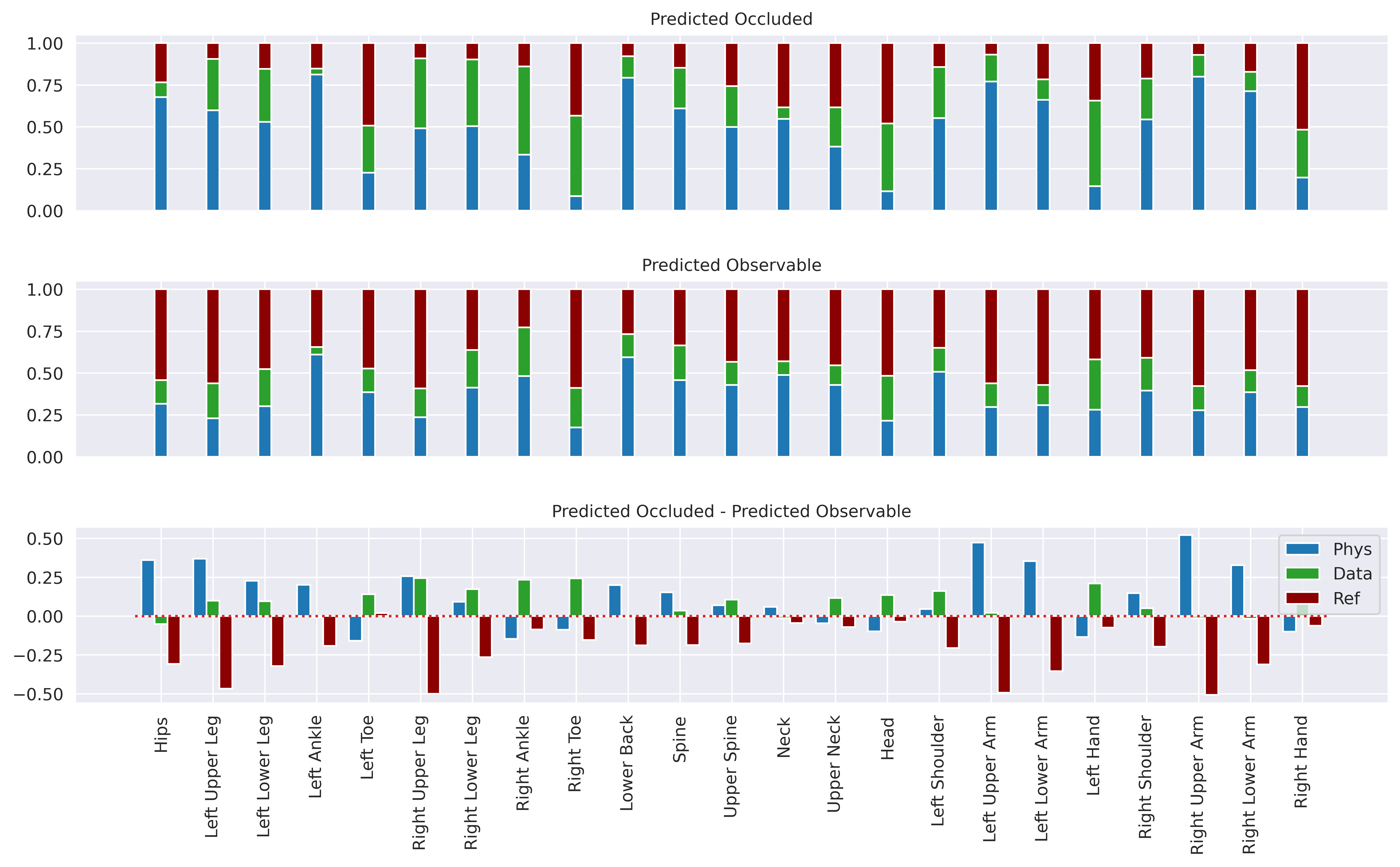}
    \caption[Per-joint fusion weights conditioned on the ``OR'' reliability
    proxy.]{Per-joint average fusion weights for the Ref-last
    (\texttt{-r}) variant conditioned on the binary ``OR'' proxy defined
    from the instantaneous reliability estimate and defiance state:
    ``Predicted Occluded'' (top) and ``Predicted Observable'' (middle).
    The difference is shown at the bottom, where positive values indicate
    an increased contribution in the predicted-occluded condition.}
    \label{fig:real_mix_occobs}
\end{figure}

Numerical value of the classifiers performance (computed with the balanced accuracy formula to consider class imbalance) is reported in Table~\ref{tab:classifiers}. The ``OR'' rule consistently achieves the highest balanced accuracy. All three rules nevertheless exhibit similar joint-wise patterns: performance is near chance for the root, moderate for axial joints,
and high for most limb joints. Since most errors for axial joints are FN, we may have a partial explanation to the lower weight mixing variations \wrt the occlusion scenario for these joints (\cf Figure~\ref{fig:mix_occobs}). In fact, Figure~\ref{fig:real_mix_occobs} shows the branches behaviour depending on the predicted class (estimated with ``OR'') and we can note a) an overall larger prevalence of the data branch under ``Predicted Occlusion'' and, b) a much more dynamic Hips, although c) the axial joints still show less behavioural shift.
Of course, such classification although interpretable and informative will not explain the end results in their entirety. In part because, as introduced previously, these signals are further processed by small MLPs (thus biased by the training set) to \eg, produce the weight handling the correction amplitude bottleneck. %
Overall, classification performance is highly joint-dependent: near-chance for the root, moderate for axial joints and high for limb joints.

\begin{figure}[t]
    \centering
    \includegraphics[width=1\linewidth]{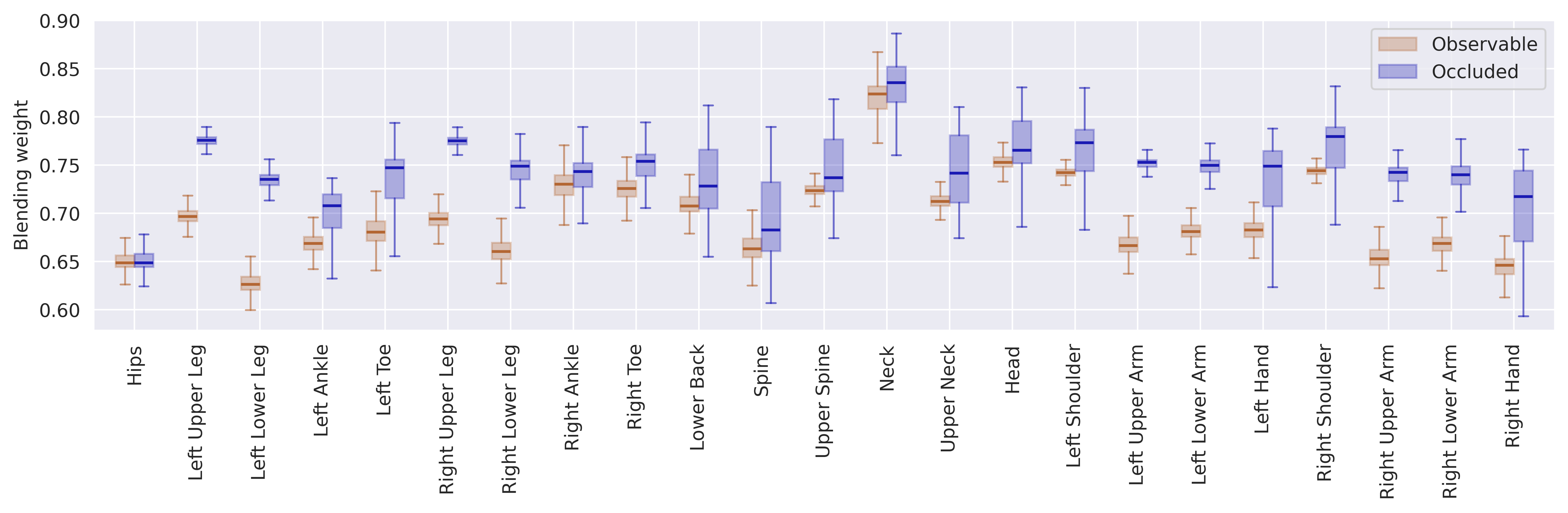}
    \caption{Geometry-aware PCAF blending weight under observable and occluded conditions. Higher value means that the PCAF-like mechanism has more impact on the end result and, as such, is more conservative.}
    \label{fig:pj_pcafw}
\end{figure}

\para{Geodesic PCAF blending}
One of the most informative signals for the end result correction is the blending weight of the geometry-aware PCAF. Specifically, based on the defiance buffer and occlusion probability regressed by the network, we compute a weight that blends the (constrained) output of the PCAF-like process and an unconstrained one. Theoretically, this weight is expected to increase with the probability of being under occlusion to enable a decrease of the correction amplitude. First, we present in Table~\ref{tab:pcafw_agg} its mean and standard deviation depending on the (actual) occlusion scenario and the variant used. We can observe that a) the weights are overall close to a specific value across joints and b) they increase when under occlusion. In particular, The \texttt{-r} variant exhibits substantially less across-joint variation than the other variants, suggesting that the learned modulation also depends on variant-specific fusion dynamics.
A more explicit distribution of the geometrical PCAF weights is presented in Figure~\ref{fig:pj_pcafw} (Ref-last variant). Overall, between Observable and Occluded, the weight distribution either gets closer to 1, or spans a wider range which may indicate a more dynamic behaviour. In particular, the legs and arms have clearly distinct regimes between both occlusion scenarios. In contrast, Hips shows little change, which is consistent with its weaker reliability classification.

\begin{table}[t]
\caption{Aggregated ($mean \pm std$ across joints) PCAF-like weight under observable and occluded conditions. Higher value represents a stronger constraint on the update correction amplitude and thus a more conservative frame.}
\label{tab:pcafw_agg}
\centering
\small
\setlength{\tabcolsep}{8pt}
\begin{tabular}{lccc}
\toprule
Scenario & Ref-last (\texttt{-r}) & Data-last (\texttt{-d}) & Phys-last (\texttt{-p}) \\
\midrule
Obs & $0.70 \pm 0.04$ & $0.65 \pm 0.15$ & $0.65 \pm 0.17$ \\
Occ & $0.75 \pm 0.03$ & $0.68 \pm 0.13$ & $0.69 \pm 0.15$ \\
\bottomrule
\end{tabular}
\end{table}

\begin{figure}[t]
    \centering
    \includegraphics[width=1\linewidth]{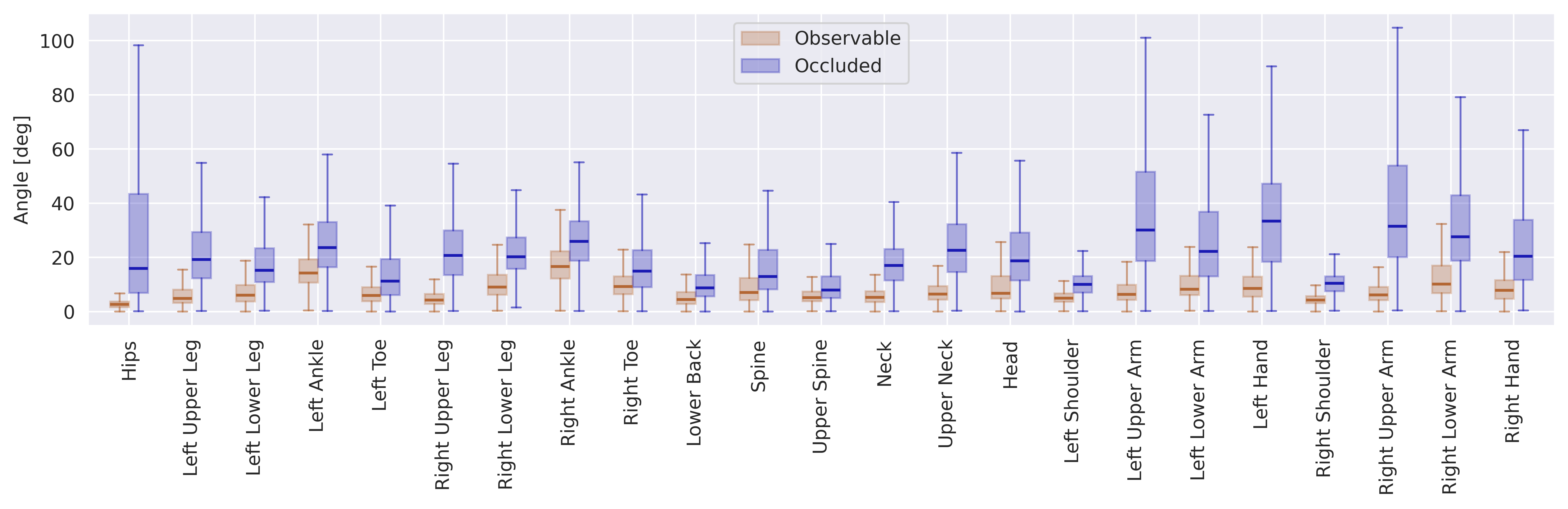}
    \caption{Local angular error under observable and occluded conditions.}
    \label{fig:supp_pjlae}
\end{figure}

\para{Per-joint angular error}
We conclude this internal analysis by examining the per-joint local
angular error of the Ref-last variant in
Figure~\ref{fig:supp_pjlae}. The results reveal no direct correspondence
between reliability classification and reconstruction difficulty.
For instance, arm joints combine relatively accurate reliability
classification with substantial angular errors, whereas the lower back
exhibits weaker classification but comparatively small reconstruction
error.

This indicates that occlusion state is only one factor governing
completion difficulty; motion amplitude, multimodality, kinematic role,
and the available contextual cues also vary substantially across joints.
Consequently, this reliability analysis should be interpreted as a
characterization of one internal mechanism rather than as a complete
explanation of the final reconstruction error.

